\documentclass{aastex63}

\submitjournal{ApJ - \today}

\shorttitle{The HAWC AGN survey}
\shortauthors{The HAWC Collaboration}

\begin{document}

\title{A survey of active galaxies at TeV photon energies with the HAWC gamma-ray observatory}

\correspondingauthor{Alberto Carrami\~nana}  \email{alberto@inaoep.mx}
\correspondingauthor{Sara Couti\~no de Le\'on}  \email{sara@inaoep.mx}
\correspondingauthor{Daniel Rosa-Gonz\'alez}  \email{danrosa@inaoep.mx}
\correspondingauthor{Anna Lia Longinotti}  \email{annalia@inaoep.mx}

\author[0000-0003-0197-5646]{A.~Albert}
 \affiliation{Physics Division, Los Alamos National Laboratory, Los Alamos, NM, USA}
\author{C.~Alvarez}
 \affiliation{Universidad Aut\'onoma de Chiapas, Tuxtla Guti\'errez, Chiapas, M\'exico}
\author{J.R. Angeles~Camacho}
 \affiliation{Instituto de F\'isica, Universidad Nacional Aut\'onoma de M\'exico, Ciudad de M\'exico, M\'exico}
\author{J.C.~Arteaga-Vel\'azquez}
 \affiliation{Universidad Michoacana de San Nicol\'as de Hidalgo, Morelia, M\'exico}
\author{K.P.~Arunbabu}
 \affiliation{Instituto de Geof\'isica, Universidad Nacional Aut\'onoma de M\'exico, Ciudad de M\'exico, M\'exico}
\author{D. Avila~Rojas}
 \affiliation{Instituto de F\'isica, Universidad Nacional Aut\'onoma de M\'exico, Ciudad de M\'exico, M\'exico}
\author[0000-0002-2084-5049]{H.A.~Ayala~Solares}
 \affiliation{Department of Physics, Pennsylvania State University, University Park, PA, USA}
\author[0000-0003-0477-1614]{V.~Baghmanyan}
 \affiliation{Institute of Nuclear Physics Polish Academy of Sciences, PL-31342 IFJ-PAN, Krakow, Poland}
\author[0000-0002-1114-2640]{E.~Belmont-Moreno}
 \affiliation{Instituto de F\'isica, Universidad Nacional Aut\'onoma de M\'exico, Ciudad de M\'exico, M\'exico}
\author[0000-0001-5537-4710]{S.Y.~BenZvi}
\affiliation{Department of Physics \& Astronomy, University of Rochester, Rochester, NY , USA} 
\author[0000-0002-5493-6344]{C.~Brisbois}
 \affiliation{Department of Physics, University of Maryland, College Park, MD, USA}
\author[0000-0002-4042-3855]{K.S.~Caballero-Mora}
 \affiliation{Universidad Aut\'onoma de Chiapas, Tuxtla Guti\'errez, Chiapas, M\'exico}
\author[0000-0003-2158-2292]{T.~Capistr\'an}
 \affiliation{Instituto Nacional de Astrof\'isica, \'Optica y Electr\'onica, Tonantzintla, Puebla, M\'exico}
\author[0000-0002-8553-3302]{A.~Carrami\~{n}ana}
 \affiliation{Instituto Nacional de Astrof\'isica, \'Optica y Electr\'onica, Tonantzintla, Puebla, M\'exico}
\author[0000-0002-6144-9122]{S.~Casanova}
 \affiliation{Institute of Nuclear Physics Polish Academy of Sciences, PL-31342 IFJ-PAN, Krakow, Poland}
\author[0000-0002-7607-9582]{U.~Cotti}
 \affiliation{Universidad Michoacana de San Nicol\'as de Hidalgo, Morelia, M\'exico}
\author[0000-0002-1132-871X]{J.~Cotzomi}
\affiliation{Facultad de Ciencias F\'isico Matem\'aticas, Benem\'erita Universidad Aut\'onoma de Puebla, Puebla, M\'exico}
\author[0000-0002-7747-754X]{S.~Couti\~{n}o~de~Le\'on}
 \affiliation{Instituto Nacional de Astrof\'isica, \'Optica y Electr\'onica, Tonantzintla, Puebla, M\'exico}
\author[0000-0001-9643-4134]{E.~De~la~Fuente}
 \affiliation{Departamento de F\'isica, Centro Universitario de Ciencias Exactas e Ingenier\'ias, Universidad de Guadalajara, Guadalajara, M\'exico}
\author[0000-0001-8451-7450]{B.L.~Dingus}
 \affiliation{Physics Division, Los Alamos National Laboratory, Los Alamos, NM, USA}
\author[0000-0002-2987-9691]{M.A.~DuVernois}
 \affiliation{Department of Physics, University of Wisconsin-Madison, Madison, WI, USA}
\author{M.~Durocher}
 \affiliation{Physics Division, Los Alamos National Laboratory, Los Alamos, NM, USA}
\author[0000-0002-0087-0693]{J.C.~D\'iaz-V\'elez}
 \affiliation{Departamento de F\'isica, Centro Universitario de Ciencias Exactas e Ingenier\'ias, Universidad de Guadalajara, Guadalajara, M\'exico}
\author[0000-0001-5737-1820]{K.~Engel}
\affiliation{Department of Physics, University of Maryland, College Park, MD, USA} 
\author[0000-0001-7074-1726]{C.~Espinoza}
 \affiliation{Instituto de F\'isica, Universidad Nacional Aut\'onoma de M\'exico, Ciudad de M\'exico, M\'exico}
\author{K.L.~Fan}
 \affiliation{Department of Physics, University of Maryland, College Park, MD, USA}
\author{M.~Fern\'andez~Alonso}
 \affiliation{Department of Physics, Pennsylvania State University, University Park, PA, USA}
\author[0000-0002-0794-8780]{H.~Fleischhack}
 \affiliation{Department of Physics, Michigan Technological University, Houghton, MI, USA}
\author[0000-0002-0173-6453]{N.~Fraija}
 \affiliation{Instituto de Astronom\'ia, Universidad Nacional Aut\'onoma de M\'exico, Ciudad de M\'exico, M\'exico}
\author{A.~Galv\'an-G\'amez}
 \affiliation{Instituto de Astronom\'ia, Universidad Nacional Aut\'onoma de M\'exico, Ciudad de M\'exico, M\'exico}
\author{D.~Garc\'ia}
 \affiliation{Instituto de F\'isica, Universidad Nacional Aut\'onoma de M\'exico, Ciudad de M\'exico, M\'exico}
\author[0000-0002-4188-5584]{J.A.~Garc\'ia-Gonz\'alez}
 \affiliation{Instituto de F\'isica, Universidad Nacional Aut\'onoma de M\'exico, Ciudad de M\'exico, M\'exico}
\author[0000-0003-1122-4168]{F.~Garfias}
 \affiliation{Instituto de Astronom\'ia, Universidad Nacional Aut\'onoma de M\'exico, Ciudad de M\'exico, M\'exico}
\author[0000-0002-5209-5641]{M.M.~Gonz\'alez}
 \affiliation{Instituto de Astronom\'ia, Universidad Nacional Aut\'onoma de M\'exico, Ciudad de M\'exico, M\'exico}
\author[0000-0002-9790-1299]{J.A.~Goodman}
 \affiliation{Department of Physics, University of Maryland, College Park, MD, USA}
\author[0000-0001-9844-2648]{J.P.~Harding}
 \affiliation{Physics Division, Los Alamos National Laboratory, Los Alamos, NM, USA}
\author[0000-0002-2565-8365]{S.~Hern\'andez}
 \affiliation{Instituto de F\'isica, Universidad Nacional Aut\'onoma de M\'exico, Ciudad de M\'exico, M\'exico}
\author{B.~Hona}
 \affiliation{Department of Physics, Michigan Technological University, Houghton, MI, USA}
\author[0000-0002-3808-4639]{D.~Huang}
 \affiliation{Department of Physics, Michigan Technological University, Houghton, MI, USA}
\author[0000-0002-5527-7141]{F.~Hueyotl-Zahuantitla}
 \affiliation{Universidad Aut\'onoma de Chiapas, Tuxtla Guti\'errez, Chiapas, M\'exico}
\author{P.~H\"untemeyer}
 \affiliation{Department of Physics, Michigan Technological University, Houghton, MI, USA}
\author[0000-0001-5811-5167]{A.~Iriarte}
 \affiliation{Instituto de Astronom\'ia, Universidad Nacional Aut\'onoma de M\'exico, Ciudad de M\'exico, M\'exico}
\author[0000-0002-6738-9351]{A.~Jardin-Blicq}
\affiliation{Max-Planck Institute for Nuclear Physics, 69117 Heidelberg, Germany}
\affiliation{Department of Physics, Faculty of Science, Chulalongkorn University, 254 Phayathai Road,Pathumwan, Bangkok 10330, Thailand}
\affiliation{National Astronomical Research Institute of Thailand (Public Organization), Don Kaeo, MaeRim, Chiang Mai 50180, Thailand}
\author[0000-0003-4467-3621]{V.~Joshi}
 \affiliation{Erlangen Centre for Astroparticle Physics, Friedrich-Alexander-Universit\"at Erlangen-N\"urnberg, Erlangen, Germany}
 \author[0000-0003-4785-0101]{D.~Kieda}
\affiliation{Department of Physics and Astronomy, University of Utah, Salt Lake City, UT, USA}
\author{G.J.~Kunde}
\affiliation{Physics Division, Los Alamos National Laboratory, Los Alamos, NM, USA}
\author[0000-0001-6336-5291]{A.~Lara}
\affiliation{Instituto de Geof\'isica, Universidad Nacional Aut\'onoma de M\'exico, Ciudad de M\'exico, M\'exico}
\author[0000-0002-2467-5673]{W.H.~Lee}
\affiliation{Instituto de Astronom\'ia, Universidad Nacional Aut\'onoma de M\'exico, Ciudad de M\'exico, M\'exico}
\author[0000-0001-5516-4975]{H.~Le\'on~Vargas}
 \affiliation{Instituto de F\'isica, Universidad Nacional Aut\'onoma de M\'exico, Ciudad de M\'exico, M\'exico}
 \author[0000-0003-2696-947X]{J.T.~Linnemann}
\affiliation{Department of Physics and Astronomy, Michigan State University, East Lansing, MI, USA}
\author[0000-0001-8825-3624]{A.L.~Longinotti}
 \affiliation{Instituto Nacional de Astrof\'isica, \'Optica y Electr\'onica, Tonantzintla, Puebla, M\'exico}
\author[0000-0003-2810-4867]{G.~Luis-Raya}
 \affiliation{Universidad Polit\'ecnica de Pachuca, Pachuca, Hidalgo, M\'exico}
\author{J.~Lundeen}
 \affiliation{Department of Physics and Astronomy, Michigan State University, East Lansing, MI, USA}
\author[0000-0001-8088-400X]{K.~Malone}
 \affiliation{Physics Division, Los Alamos National Laboratory, Los Alamos, NM, USA}
\author[0000-0001-9052-856X]{O.~Mart\'inez}
\affiliation{Facultad de Ciencias F\'isico Matem\'aticas, Benem\'erita Universidad Aut\'onoma de Puebla, Puebla, M\'exico}
\author[0000-0001-9035-1290]{I.~Martinez-Castellanos}
\affiliation{Department of Physics, University of Maryland, College Park, MD, USA}
\author[0000-0002-2824-3544]{J.~Mart\'inez-Castro}
 \affiliation{Centro de Investigaci\'on en Computaci\'on, Instituto Polit\'ecnico Nacional, Ciudad de M\'exico, M\'exico}
\author[0000-0002-2610-863X]{J.A.~Matthews}
 \affiliation{Dept of Physics and Astronomy, University of New Mexico, Albuquerque, NM, USA}
\author[0000-0002-8390-9011]{P.~Miranda-Romagnoli}
 \affiliation{Universidad Aut\'onoma del Estado de Hidalgo, Pachuca, M\'exico}
 \author{J.A.~Morales-Soto}
\affiliation{Universidad Michoacana de San Nicol\'as de Hidalgo, Morelia, M\'exico}
\author[0000-0002-1114-2640]{E.~Moreno}
 \affiliation{Facultad de Ciencias F\'isico Matem\'aticas, Benem\'erita Universidad Aut\'onoma de Puebla, Puebla, M\'exico}
\author[0000-0002-7675-4656]{M.~Mostaf\'a}
 \affiliation{Department of Physics, Pennsylvania State University, University Park, PA, USA}
\author[0000-0003-0587-4324]{A.~Nayerhoda}
 \affiliation{Institute of Nuclear Physics Polish Academy of Sciences, PL-31342 IFJ-PAN, Krakow, Poland}
\author[0000-0003-1059-8731]{L.~Nellen}
 \affiliation{Instituto de Ciencias Nucleares, Universidad Nacional Aut\'onoma de M\'exico, Ciudad de M\'exico, M\'exico}
\author[0000-0001-9428-7572]{M.~Newbold}
 \affiliation{Department of Physics and Astronomy, University of Utah, Salt Lake City, UT, USA}
\author[0000-0002-0479-2311]{M.U.~Nisa}
 \affiliation{Department of Physics and Astronomy, Michigan State University, East Lansing, MI, USA}
\author[0000-0001-7099-108X]{R.~Noriega-Papaqui}
 \affiliation{Universidad Aut\'onoma del Estado de Hidalgo, Pachuca, M\'exico}
\author{L.~Olivera-Nieto}
\affiliation{Max-Planck Institute for Nuclear Physics, 69117 Heidelberg, Germany}
\author{A.~Peisker}
 \affiliation{Department of Physics and Astronomy, Michigan State University, East Lansing, MI, USA}
\author[0000-0001-5998-4938]{E.G.~P\'erez-P\'erez}
 \affiliation{Universidad Polit\'ecnica de Pachuca, Pachuca, Hidalgo, M\'exico}
\author[0000-0002-6524-9769]{C.D.~Rho}
 \affiliation{University of Seoul, Seoul, Rep. of Korea}
\author[0000-0003-1327-0838]{D.~Rosa-Gonz\'alez}
 \affiliation{Instituto Nacional de Astrof\'isica, \'Optica y Electr\'onica, Tonantzintla, Puebla, M\'exico}
\author[0000-0001-6939-7825]{E.~Ruiz-Velasco}
\affiliation{Max-Planck Institute for Nuclear Physics, 69117 Heidelberg, Germany}
\author{H.~Salazar}
 \affiliation{Facultad de Ciencias F\'isico Matem\'aticas, Benem\'erita Universidad Aut\'onoma de Puebla, Puebla, M\'exico}
\author[0000-0002-8610-8703]{F.~Salesa~Greus}
\affiliation{Institute of Nuclear Physics Polish Academy of Sciences, PL-31342 IFJ-PAN, Krakow, Poland}
\affiliation{Instituto de F\'isica Corpuscular, CSIC, Universidad de Valencia, E-46980, Paterna, Valencia, Espa\~na}
\author[0000-0001-6079-2722]{A.~Sandoval}
 \affiliation{Instituto de F\'isica, Universidad Nacional Aut\'onoma de M\'exico, Ciudad de M\'exico, M\'exico}
\author[0000-0001-8644-4734]{M.~Schneider}
 \affiliation{Department of Physics, University of Maryland, College Park, MD, USA}
\author[0000-0002-8999-9249]{H.~Schoorlemmer}
\affiliation{Max-Planck Institute for Nuclear Physics, 69117 Heidelberg, Germany}
\author[0000-0002-1012-0431]{A.J.~Smith}
\affiliation{Department of Physics, University of Maryland, College Park, MD, USA}
\author[0000-0002-1492-0380]{R.W.~Springer}
 \affiliation{Department of Physics and Astronomy, University of Utah, Salt Lake City, UT, USA}
\author[0000-0001-9725-1479]{K.~Tollefson}
 \affiliation{Department of Physics and Astronomy, Michigan State University, East Lansing, MI, USA}
\author[0000-0002-1689-3945]{I.~Torres}
 \affiliation{Instituto Nacional de Astrof\'isica, \'Optica y Electr\'onica, Tonantzintla, Puebla, M\'exico}
\author{R.~Torres-Escobedo}
 \affiliation{Departamento de F\'isica, Centro Universitario de Ciencias Exactas e Ingenier\'ias, Universidad de Guadalajara, Guadalajara, M\'exico}
\author{F.~Ure\~na-Mena}
 \affiliation{Instituto Nacional de Astrof\'isica, \'Optica y Electr\'onica, Tonantzintla, Puebla, M\'exico}
\author[0000-0001-6876-2800]{L.~Villase\~{n}or}
 \affiliation{Facultad de Ciencias F\'isico Matem\'aticas, Benem\'erita Universidad Aut\'onoma de Puebla, Puebla, M\'exico}
\author{T.~Weisgarber}
 \affiliation{Department of Chemistry and Physics, California University of Pennsylvania, California, Pennsylvania, USA}
 \author{E.~Willox}
\affiliation{Department of Physics, University of Maryland, College Park, MD, USA} 
\author[0000-0001-9976-2387]{A.~Zepeda}
 \affiliation{Departamento de F\'isica, Centro de Investigaci\'on y de Estudios Avanzados del IPN, Ciudad de M\'exico, M\'exico}
\author[0000-0003-0513-3841]{H.~Zhou}
 \affiliation{Tsung-Dao Lee Institute \& School of Physics and Astronomy, Shanghai Jiao Tong University, Shanghai, China}
\author[0000-0002-8528-9573]{C. de~Le\'on}
 \affiliation{Universidad Michoacana de San Nicol\'as de Hidalgo, Morelia, M\'exico}
\collaboration{90}{The HAWC collaboration}


\begin{abstract}
The High Altitude Water Cherenkov Gamma-Ray Observatory (HAWC) continuously detects TeV photons and particles within its large field-of-view, accumulating every day a deeper exposure of two thirds of the sky. We analyzed 1523~days of HAWC live data acquired over four and a half years, in a follow-up analysis of {138} nearby ($z<0.3$) active galactic nuclei from the {\em Fermi} 3FHL catalog culminating within $40^\circ$ of the zenith at Sierra Negra, the HAWC site. This search for persistent TeV emission used a maximum-likelihood analysis assuming intrinsic power-law spectra attenuated by pair production of gamma-ray photons with the extragalactic background light. HAWC clearly detects persistent emission from Mkn~421 and Mkn~501, the two brightest blazars in the TeV sky, at 65$\sigma$ and 17$\sigma$ level, respectively. {Weaker evidence for long-term emission is found for three other known very-high energy emitters:} the radiogalaxy M87 and the BL Lac objects VER~J0521+211 and 1ES~1215+303, the later two at $z\sim 0.1$. We find evidence for collective emission from the set of 30 previously reported very high-energy sources that excludes Mkn~421 and Mkn~501 with a random probability $\sim 10^{-5}$. Upper limits are presented for the sample under the power-law assumption and in the predefined (0.5-2.0), (2.0-8.0) and (8.0-32.0) TeV energy intervals.
\end{abstract}



\section{Introduction \label{sec:intro}}
The nuclei of active galaxies are remarkable in possessing observed bolometric luminosities up to $10^{48}-10^{49}\,\rm erg/s$, surpassing the energy output of their host galaxies, while being unresolved down to the smallest physical scales observable. The stringent requirements of extreme energy outputs in restricted volumes, together with rapid variability and the presence of powerful relativistic jets, led early on to modeling active galactic nuclei (AGN) as accreting supermassive black holes of masses up to $10^{9}~\rm M_{\odot}$ and accretion rates exceeding $1~\rm M_{\odot}/year$~\citep{1963MNRAS.125..169H,1999qagn.book.....K}. The geometry of an inner accretion disk and an outer dusty torus surrounding the black hole allows for a qualitative view of the different types of AGN in terms of the orientation of the line-of-sight relative to the disk and torus axes aligned with the black hole rotation axis~\citep{1993ARA&A..31..473A}.  Anisotropic emission causes over-estimates of the AGN luminosity for privileged lines-of sight. In standard AGN scenarios, jets can be powered either by the inner regions of the radiation dominated accretion disk~\citep{1982MNRAS.199..883B,2015SSRv..191..441H}, or by the rapid rotation of the black hole~\citep{1977MNRAS.179..433B,2011MNRAS.418L..79T}. 

AGN constitute the most common type of GeV $\gamma$-ray source in the sky. Most of the objects detected with the Large Area Telescope (LAT) on board {\em Fermi} are extragalactic, and the vast majority of them are blazars, either BL Lac objects or flat-spectrum radio quasars~\citep{2015ApJS..218...23A,2017ApJS..232...18A}. Relativistic models considering the alignment of AGN jets with our line of sight can account, in most cases, for the high energetics and rapid variability observed in $\gamma$-ray emission up to TeV energies~\citep{1993ApJ...416..458D,2018Galax...6...68L}. The detection of radiogalaxies and Seyfert galaxies with the {\em Fermi}-LAT and Imaging Atmospheric Cherenkov Telescopes (IACTs) provides further insight into the jet powered view of AGN under a unified scheme, as probes of off-axis $\gamma$-ray emission~\citep{2018Galax...6..116R}.

For over three decades AGN have been suspected to be the sources of ultra-high-energy cosmic rays: acceleration may occur either in the central engine, the relativistic jet, or in distant radio lobes~\citep{1984ARA&A..22..425H,2011ARA&A..49..119K}. The coincidence of the 290~TeV neutrino event IceCube-170922A with the $\gamma$-ray emitting BL Lac TXS~0506+056 provided fresh observational support~\citep{2018Sci...361..147I}. Observations at the highest photon energies are important to characterize the extreme energetics of AGN. While space borne instruments conduct deep and wide field-of-view observations leading to all-sky surveys in the high-energy range (HE; 0.1-100~GeV), most of our knowledge in the very-high energy (VHE; $\gtrsim 100\,\rm GeV$) regime comes from pointed observations with IACTs, nowadays able to detect individual sources with fluxes greater than 1/1000 that of the Crab Nebula. IACT observatories rely on performing deep, but sparse, follow-up observations of active objects, adequate selections of targets and surveying regions of particular interest. They are limited by their relatively small fields-of-view, with the most extensive survey performed with these instruments to date, the dedicated H.E.S.S. Galactic Plane Survey, covering $\sim$0.3~steradians in the course of a decade~\citep{2018A&A...612A...1H}.

Compensating their lower instantaneous sensitivity with steradian fields-of-view, high altitude air shower arrays are now able to perform unbiased continuous monitoring of known AGN, in particular Markarian~421 and Markarian~501~\citep{2017ApJ...841..100A}. This preludes their potential to conduct large surveys with sufficient depth to detect extragalactic TeV sources. The main hurdle for reaching the extragalactic sky is the access to the lowest photon energies, as our view of the TeV sky is impaired by extragalactic background light (EBL). Photon-photon pair production of TeV $\gamma$ rays with EBL photons sets a physical limit on how far in distance and in spectral range extragalactic source can be observed~\citep{1998ApJ...493..547S}. Pair production proceeds efficiently just above its kinematic threshold, making TeV $\gamma$ rays prone to interact with infrared radiation.

This paper presents an AGN follow-up survey performed with four and a half years of full-operations data from the High Altitude Water Cherenkov (HAWC) Gamma-Ray Observatory, building on the preliminary release of this work with a somewhat smaller HAWC dataset~\citep{2019arXiv190806831C}. The HAWC time-integrated data cover $\sim 60\%$ of the sky, extending the {\em Fermi}-LAT all-sky survey to a search for persistent TeV $\gamma$-ray emission. The HAWC survey encompasses all AGN in the Third Catalog of Hard {\em Fermi}-LAT sources (3FHL)  accesible from the HAWC site with a redshift $z\leq 0.3$, of which about 20\% have been previously reported in the VHE range through IACT pointed observations in different states of activity. 
The paper is organized as follows: 
\S\ref{sec:hawcgro} describes the HAWC observatory, its data and standard analysis; 
\S\ref{sec:ebl} provides a brief review of the EBL and its effect on TeV spectra; 
\S\ref{sec:sample} presents the sample of AGN drawn from the 3FHL catalog, with due considerations of ground-based IACT observations; 
\S\ref{sec:3fhl-follow-up} discusses the dedicated follow-up study of the sample, leading to the summary and conclusions in \S\ref{sec:summary}.

\section{The HAWC Gamma-Ray Observatory \label{sec:hawcgro}}
HAWC is a wide field-of-view TeV $\gamma$-ray observatory optimized for surveying cosmic high-energy sources. It is located inside the Parque Nacional Pico de Orizaba, in the Mexican state of Puebla. The HAWC array occupies a relatively flat area of the Volc\'an Sierra Negra mountain, at an altitude of 4100~meters, centered at geographical latitude $18.995^\circ\rm N$ and longitude $97.308^\circ\rm W$. HAWC has achieved a $\gtrsim 95\%$ duty cycle, allowing it to survey two thirds of the sky every sidereal day with sufficient depth to detect the Crab Nebula at the $5\sigma$ level~\citep{2017ApJ...843...39A}. This study improves on the 2HWC survey performed with the first year and a half of data, that allowed the detection of 39 sources of TeV $\gamma$ rays~\citep{2017ApJ...843...40A}. The most recent HAWC all-sky survey, 3HWC, uses the same data as in here. It contains 65~TeV $\gamma$-ray sources, most of them along the Galactic Plane, including the $189\sigma$ detection of the Crab Nebula~\citep{2020arXiv200708582A}.

\subsection{The HAWC detector}
HAWC is an extensive air shower (EAS) array sampling in detail secondary particles produced by primary cosmic rays in the upper atmosphere. HAWC data analysis can distinguish between hadronic and $\gamma$-ray induced cascades through their different charge distributions at the ground. The observatory consists of a dense array of 300 large water Cherenkov detectors (WCDs) covering collectively a physical area larger than 22,000~m$^2$. Each WCD is a cylindrical tank of 7.3~m diameter and 5~m height, filled with 180~m$^3$ of water and instrumented with four upward-facing photomultiplier tubes (PMTs) at its base. The signals from the 1200 PMT channels are brought to the data acquisition system located near the center of the array to be processed in real time. HAWC has been in full-operations since its inauguration on March 20, 2015, after two years of gathering data with a partial array configuration~\citep{2016ApJ...817....3A}. Further details about the observatory can be found in \citet{2017ApJ...843...39A}.

\subsection{HAWC data and standard analysis \label{subsec:analysis}}
The HAWC array records about 25,000 events per second, the vast majority caused by hadronic cosmic rays. Each data record contains the particle arrival timing and deposited charge on each of the 1200 PMTs which are used to locate the event in the sky and to perform photon/hadron discrimination. The analysis presented here follows the validation observation of the Crab Nebula, both in the  $\gamma$-hadron cuts used and in partitioning the data in nine bins according to the fraction of channels hit, as indicated in Table~2 of \citet{2017ApJ...843...39A}. The bin number provides a coarse measure of the primary energy, with an important overlap in the energy distributions of different bins due to the fluctuations inherent in the development of particle cascades. Lower bins relate to lower energies, and the spatial resolution improves with increasing bin number, with the detailed detector response depending on the spectrum of the source and its declination. {The lowest bin used in this analysis, ${\cal B}=1$, has peak sensitivity around $0.5~\rm TeV$ for a source with a power-law spectrum of index 2.63 culminating at the zenith, as the Crab Nebula.}

The HAWC sky surveys have a mean photon energy of about 7~TeV and one-year sensitivity between 50 and 100~mCrab~\citep{2017ApJ...843...40A}. HAWC data analysis is based on computing the likelihood ratio of a source+background to a background only models, given by the test statistic,
\begin{equation}
{TS} = 2 \ln\left\{ {\cal L}(S+B) \over {\cal L}(B) \right\} , \label{ts}
\end{equation}
where ${\cal L}(A)$ is the likelihood of model $A$, given by the product of the probability density function computed at each point of the region of interest. Given a $TS$ value, its statistical significance can be approximated by $s=\pm\sqrt{TS}$, with the sign indicating {an excess or deficit} relative to the background. For all-sky surveys, like 2HWC, the analysis is performed by optimizing $TS$ on every pixel of a $N_{\rm side}=1024$ HEALPix grid model of the observable sky~\citep{2005ApJ...622..759G}. The source model generally consists of either a point source or an extended source hypothesis following a simple power-law spectrum of fixed spectral index, with free  normalization. Joint normalization and spectral index optimizations are then performed to further characterize detected sources. The analysis presented here is performed similarly to that of the 2HWC, although on pre-defined sky locations and including the attenuation of TeV photons caused by their interaction with extragalactic background light. We use 1523~days of live data acquired between November 26, 2014, and June 3, 2019. The live data comprises 92.3\% of the total time span. The data deficit is due mostly to quality cuts and run losses during bad weather. 

The comparison of HAWC and {\em Fermi}-LAT data requires the consideration of the respective systematic uncertainties of each experiment. HAWC fluxes presented here have an estimated 15\% systematic uncertainty (\S\ref{subsec:analysis}); those in the 3FHL catalog are quoted to have uncertainties of 9\% in the 150-500~GeV band, and 15\% in 0.5-2.0~TeV~\citep{2017ApJS..232...18A}. 

\section{Photon-photon attenuation by extragalactic background light} \label{sec:ebl}
The astrophysical relevance of the $\gamma\gamma\to {\rm e}^{+}{\rm e}^{-}$ process as an absorption mechanism for distant sources was pointed out by~\citet{1966PhRvL..16..252G,1967PhRv..155.1404G,1967PhRv..155.1408G}, first in consideration of the cosmic microwave background, and later for more generic backgrounds. Photon-photon pair production is described by the cross section $\sigma_{\gamma\gamma}=\pi r_e^{2}\,\psi(\omega)$, with $r_e$ the classical electron radius, and $\psi$ an analytical function of \mbox{$\omega=\sqrt{E_1 E_2 (1-\mu)/2}$},~the energy of each photon in the center of momentum frame, a relativistic invariant given by the product of the energies of the two photons in an arbitrary frame, $E_{1}=E_{\gamma},\, E_{2}=h\nu$, and $\mu=\cos\theta$, where $\theta$ is their interaction angle. Pair-production requires $\omega\geq m_ec^{2}$, with the cross section maximized at $\omega\approx 1.4~m_ec^{2}\,\Rightarrow\, E_\gamma h\nu\approx 2 (m_e c^{2})^{2} \approx 0.5~\rm TeV\cdot eV$. Hence, 1~TeV $\gamma$ rays are prone to interact with near-infrared photons of 0.5~eV ($\lambda\simeq 2.5\,\mu{\rm m}$), while a 100~TeV photon is to interact with far-infrared extragalactic light, $\lambda\simeq 250\,\mu{\rm m}$.

The absorption of high-energy photons from a distant source of redshift $z$ traversing through intervening radiation is governed by the optical depth,
\begin{equation}
\tau(E_\gamma, z) =  {1\over 2}  \int_{0}^{d(z)}  \int_{0}^{\infty} \int_{-1}^{+1} \sigma_{\gamma\gamma}\left(\omega\right)\, n_\nu(\ell) \, d\mu \, d\nu\, d\ell , \label{ebl-tau}
\end{equation}
where the photon density $n_\nu$ may describe local and/or cosmic intervening radiation fields\footnote{eq.~\ref{ebl-tau} assumes an isotropic $n_\nu$; this may not describe a local radiation field.}. In the case of an evolving cosmic field, a dependence on redshift $n_\nu=n_\nu(z)$ may be introduced. The photon path is integrated using the cosmological light-travel distance $d\ell = c\,dz/(1+z)H(z)$. The probability that the $\gamma$ ray survives the journey is $\exp(-\tau)$. Given the usual functional form of $\tau(E_\gamma,z)$, the survival probability behaves close to a cutoff once $\tau=1$ is reached. In principle the opacity of the Universe to VHE $\gamma$ rays is calculated given $n_\nu(z)$. In practice measurements of the light backgrounds are difficult to perform, particularly in the infrared and far-infrared, and observations of distant $\gamma$-ray sources become relevant for constraining the spectral shape of the EBL, both in the local Universe and as a function of redshift~\citep{2019MNRAS.486.4233A}. In here we use the EBL model of~\citet{2011MNRAS.410.2556D}, which fits well observations by {\em Fermi}-LAT and IACTs.

We note that two of the sources in our sample NGC~1068 and M87, are located below the lower bound of the redshift range of most EBL models, which start at $z=0.01$. For these sources we assume that the photon density has not changed between now and then, to approximate,
\begin{equation}
\tau(E_\gamma, z) \approx \tau(E_\gamma, 0.01) \left(z/0.01\right) \, .  
\end{equation}

The analysis in here assumes intrinsic power-law spectra for the sources. We can describe the overall effect of the EBL estimating the observed integrated photon flux ($N_{obs}$; photons cm$^{-2}$s$^{-1}$) for an intrinsic spectrum of index $\alpha$, introducing the relation,
\begin{equation}
N_{obs}(\geq E_0) = \int_{E_0}^{\infty} \left(dN\over dE\right)_{intr} e^{-\tau(E,z)} dE  \quad = e^{-z/z_h} N_{intr}(\geq E_0) \, , \label{nobs-ebl}
\end{equation}
where $N_{intr}(\geq E_0)\propto E_0^{-\alpha+1}/(\alpha-1)$ is the integral of the differential intrinsic spectrum. The resulting horizon scale $z_h$ depends strongly on $E_0$ and weakly on the power-law index $\alpha$. Using the EBL model of~\citet{2011MNRAS.410.2556D}, we get $z_h=0.106$  for $\alpha=2.5$ and $E_0=0.5\,\rm TeV$, justifying the bound $z\leq 0.3$ considered for this study. The value of $z_h$ at 0.5~TeV ranges from 0.096 for $\alpha=2.0$, to 0.113 for $\alpha=3.0$; on the other hand, the dependence of $z_h$ with $E_0$ is exponential, going from $z_h=0.728$ at 0.1~TeV, to $z_h=0.068$ at $E_0=1.0\,\rm TeV$, for $\alpha=2.5$, as presented in \citet{2019arXiv190806831C}. 

{We compared three main EBL models for the case $\alpha=2.5,\; E_0=0.5\,\rm TeV$. When considering the upper and lower uncertainties of \citet{2011MNRAS.410.2556D}, we get that $z_h$ is in the interval $\left(0.098,\, 0.121\right)$, which is consistent to $z_h=0.102$ obtained with \citet{2012MNRAS.422.3189G} and $z_h=0.099$ with \citet{2017A&A...603A..34F}. These three models coincide within a $\sim 10\%$ systematic uncertainty.}

\section{Active galaxies above 10 GeV: 3FHL and TeV pointed observations  \label{sec:sample}}

\subsection{A Sample of Active Galaxies from the 3FHL Catalog \label{subsec:sample}}
The 3FHL catalog contains 1556 objects detected at photon energies between 10~GeV and 2~TeV in the first seven years of {\em Fermi} operations, from August 4, 2008, to August 2, 2015~\citep{2017ApJS..232...18A}.  Of the 3FHL entries, 79\% are identified or associated with extragalactic objects, mostly BL Lac objects and flat-spectrum radio quasars (48\% and 11\% of the 3FHL, respectively). As defined in the different LAT catalogs, an association refers to the positional coincidence of the HE $\gamma$-ray source with an object having suitable properties; while an identification requires measuring correlated variability between the $\gamma$-ray source and its associated counterpart. These criteria result in 9\% of the sources in 3FHL listed as identified and 78\% as associated, with the remaining 13\% are unassociated or unclassified. It is customary in {\em Fermi} catalogs to distinguish between identifications using upper case letters, as RDG for identified radio-galaxies, and associations using lower case letters, as rdg for an association with a radio-galaxy.

\begin{deluxetable}{lrc}
\tablecaption{Classes of AGN selected. \label{3fhl-selection}}
\tablehead{\colhead{Source class} & \colhead{Id + As} & \colhead{Total} }
\startdata
\hline
BL Lac objects (BLL + bll) & 6 + 111 & 117 \\
Blazars candidates of uncertain type (bcu) & 8 & 8 \\
Radiogalaxies (RDG + rdg) & 2 + ~~~4 & 6 \\
Flat spectrum radio quasars (FSRQ + fsrq) & 1 + ~~~5 & 6 \\
Starburst galaxies (SBG + sbg) & 1 & 1 \\
\hline
{\bf Total in sample} & 9 + 129 & {\bf 138} \\
\hline
\enddata
\tablecomments{Id = identified sources; As = associated sources}
\end{deluxetable}

The 3FHL catalog partitions its nominal wide 10~GeV-2~TeV spectral interval into five bands. The majority of AGN are detected at $TS>10$ in the two lower energy bands, 10-20~GeV and 20-50~GeV. Spectral cutoffs at energies $\lesssim 50\,\rm GeV$ are common, as it can be noticed in our sample: of the 138 AGN, 50 (14) are detected above $5\sigma$ in the LAT 50-150~GeV (150-500~GeV) intermediate band(s). Furthermore, Mkn~421 and Mkn~501 are the only two AGN studied here with $TS>25$ in the 0.5-2.0~TeV LAT band, the spectral intersection with HAWC. 

The 3FHL catalog assigns a flag to sources as follows: TeV=P, when reported at VHE energies; TeV=C for candidates for TeV detection; and TeV=N for non-reported and not candidates. TeV candidates are by definition sources undetected with VHE ground based instruments whose LAT data satisfy three conditions: (i) significance $s>3$ above 50~GeV; (ii) spectral index $< 3$; (iii) integrated photon flux $N(>50\,{\rm GeV})> 10^{-11}\,{\rm cm^{-2}s^{-1}}$. In addition, the 3FHL catalog assesses source variability through the $V_{\rm bayes}$ parameter, the number of Bayes blocks needed to model the light curve. A source with $V_{\rm bayes}=1$ is consistent with a constant flux~\citep{2017ApJS..232...18A}.

\begin{figure}
\centerline{\includegraphics[width=\hsize]{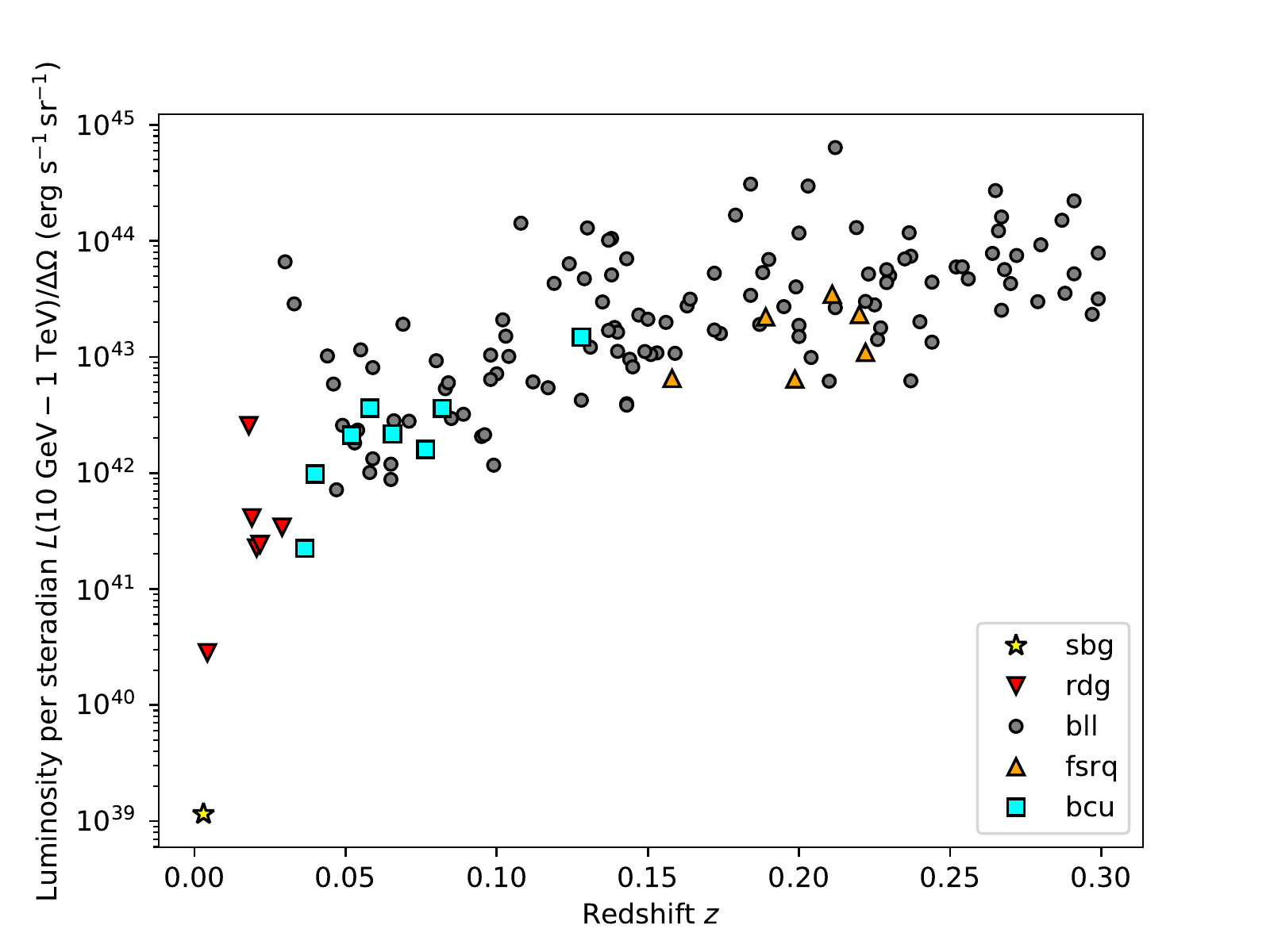}}
\caption{Luminosity as a function of redshift for the 3FHL sample studied with HAWC. Note the clear separation in redshift and luminosities between the single starburst (yellow point, the nearest object), radio-galaxies (red), intermediate distance bcu's (in blue) and the more distant group of FSRQs (orange). BL Lac objects, in grey, span most of the redshift interval and have the highest observed luminosities.  
\label{muestra-z-lumi}}
\end{figure}

We select 3FHL catalog sources identified or associated with AGN with redshifts $z\leq 0.3$ that culminate within $40^\circ$ of the zenith as viewed from the HAWC site.
For the selection of the follow-up sample we used the current version of the 3FHL catalog available at the Fermi Science Support Center\footnote{fits file gll$\_$psch$\_$v13.fits, dated July 2017 at https://fermi.gsfc.nasa.gov/ssc/.}. The sample of 138 objects is summarized in Table~\ref{3fhl-selection}. It contains 32 objects flagged TeV=P (positive VHE detections), and 32 TeV=C (candidates). The sample is grouped in five source classes, defined in the 3FHL, of distinct properties (Table~\ref{3fhl-selection}): 
\begin{itemize}
\item {\bf Starburst galaxies} are the nearest and apparently less luminous AGNs in {\em Fermi}-LAT. While prone to host active nuclei, the prevailing $\gamma$-ray emission is dominated by cosmic rays produced in star formation processes. NGC~1068 is the only 3FHL starburst inside the declination range of our selection, and the lowest redshift AGN in our sample~(Figure~\ref{muestra-z-lumi}). Intriguingly, it has recently been associated to one of the hot spots in the neutrino sky, as observed by the IceCube observatory~\citep{2020PhRvL.124e1103A}. 
\item {\bf Radiogalaxies (RDG)} are the nearest extragalactic GeV sources dominated by an active nucleus. They appear up to three orders of magnitude more luminous than starbursts~(Figure~\ref{muestra-z-lumi}). Radiogalaxies are attractive targets for HAWC as the relatively close distance translates in a reduced photon-photon attenuation, potentially allowing to sample the far-infrared portion of the EBL through observations above $\sim 10-30~\rm TeV$. Six 3FHL catalog radiogalaxies transit through the field-of-view of HAWC, with redshifts between $z=0.0042$, for M87, and $z=0.029$, for NGC~1218. Four of them have been claimed as VHE sources by IACT collaborations\footnote{3C~264 has a TeV=N flag, as its VHE detection occurred after the publication of the 3FHL catalog.}~\citep{2018Galax...6..116R}.
\item {\bf BL Lacertae objects (BLL)} constitute the majority of known GeV and VHE $\gamma$-ray emitters. They dominate the 3FHL catalog, in particular for redshifts $z\lesssim 0.7$. As expected, BL Lac objects completely dominate our sample with 6 identifications and 111 associations. They span most of the $z\leq 0.3$  range, led by Mkn~421 at $z=0.031$, the nearest and brightest 3FHL AGN.
\item {\bf Flat Spectrum Radio Quasars (FSRQ)} are the most distant and seemingly luminous blazars. Our sample includes six such sources, with redshifts ranging between $z=0.158$ to $z=0.222$. Of these, only PKS~0736+017 has been reported as a TeV source, the nearest FSRQ claimed in the VHE range so far~\citep{2017ICRC...35..627C}.
\item {\bf Blazars Candidates of Uncertain type (bcu)} are AGN poorly characterized across the electromagnetic spectrum. The 3FHL catalog contains 290 bcu's, a good fraction of them in the Southern sky, all with radio-loud associations, and 90\% of them lacking redshift measurements. Only eight such objects satisfy our selection criteria. They are relatively near objects, between $z=0.036$ and $z=0.128$~(Figure~\ref{muestra-z-lumi}). None is a VHE source, but three of them are flagged as TeV candidates.
\end{itemize}
Figure~\ref{muestra-z-lumi} shows the luminosity per solid angle unit, $d_L(z) f_e$, with $f_e$ being the (10~GeV-1~TeV) energy flux from the 3FHL catalog, and $d_L(z)$ the luminosity distance. The actual luminosity of each source depends on the unknown solid angle of emission.

{Prior to the analysis, we identified five objects in our sample close in the sky to bright 2HWC sources that could affect their analysis. We set a conservative distance threshold of 5$^\circ$, equivalent for five times the 68\% containment radius of ${\cal B}=1$, the bin where contamination by a bright source nearby is more likely. As our sample excludes by construction low Galactic latitudes, with all selected sources located at $|b|>5^\circ$, the only concerns were for:
\begin{itemize} 
\item 3FHL~J0521.7+2112 located at 3.07$^{\circ}$ from the Crab Nebula; 
\item 3FHL~J1041.7+3900, 3FHL~J1100.3+4020, and 3FHL~J1105.8+3944 located respectively at $4.51^\circ$, $2.28^\circ$ and $1.57^\circ$ from Markarian~421;
\item 3FHL~J1652.7+4024 located at $0.68^\circ$ from Markarian~501.
\end{itemize}
Only 3FHL~J0521.7+2112 is associated to a well-known VHE source, the BL Lac object VER~J0521+211. The other four sources are flagged as TeV=N. These five objects were analyzed and tested later for contamination from the bright neighbor source, as detailed for VER~J0521+211 in \S\ref{ver_j0521+211}. From the respective tests we decided to exclude 3FHL~J1105.8+3944 and 3FHL~J1652.7+4024 from our sample.}

We also revised the redshift measurements listed in the 3FHL catalog. Our sample is dominated by BL Lac objects, often with questionable redshifts. Redshifts from the 3FHL catalog were systematically collated with the SIMBAD and NED databases, with the SDSS database consulted in some particular cases. We used as references the redshift surveys of \citet{2013ApJ...764..135S} and \citet{2008ApJS..175...97H}, and in particular the dedicated survey of TeV sources performed with the 10.4m Gran Telescopio Canarias (GTC) by \citet{2017ApJ...837..144P}. While we decided to systematically use the redshifts listed in the 3FHL catalog, we list in Table~\ref{revision-redshift} sources with disputed values, for future reference.

\begin{deluxetable*}{llcccc}
\tablecaption{Selected sources with questioned redshift.\label{revision-redshift}}
\tablehead{
\colhead{3FHL name} & \colhead{Association} & \colhead{$z_{3FHL}$} & \colhead{$z_{rev}$} & \colhead{Notes} & \colhead{References}}
\startdata
3FHL~J0112.1+2245 & S2~0109+22 & 0.265 & [0.35, 0.67] & Detected up to 200~GeV & (1,2,3) \\
3FHL~J0521.7+2112 & TXS 0518+211 & 0.108 & $>0.18$ &Emission up $\gtrsim 1\,\rm TeV$.  & (4,5,6) \\
3FHL~J0650.7+2503 & 1ES~0647+250 & 0.208 & $>0.29$ &  $z\simeq 0.41$ from host properties. & (4,5,7) \\
3FHL~J1230.2+2517 & ON~246 & 0.135 & $>0.10$ & -- & (5,8) \\
3FHL~J2323.8+4210 & BZB~J2323+4210 & 0.059 & $\geq 0.267$ & TeV candidate & (5) \\
\enddata
\tablerefs{(1)\citet{2008ApJS..175...97H}; (2) \citet{2016MNRAS.458.2836P}; (3) \citet{2018MNRAS.480..879M}; (4) \citet{2013ApJ...764..135S}; (5) \citet{2017ApJ...837..144P};  (6) \citet{2015ICRC...34..864P}; 
(7) Kotilainen et al. (2011); (8) Nass et al. (1996). }
\end{deluxetable*}

\subsection{VHE Ground-Based Observations Related to Our Sample}
AGN have been extensively studied with IACTs, with evidence of VHE $\gamma$-ray emission in almost a hundred of them~\citep{2008ICRC....3.1341W,2016ARA&A..54..725M}. These observations have shown that TeV flaring is an intrinsic characteristic of blazars and radiogalaxies. IACT deep observations, able to reach down a few percent of the flux of the Crab Nebula in a single run, have identified high, medium and low states of activity in several sources. However, continuous long-term coverage of the AGN population cannot be performed with IACTs. The definition of the base level of AGN VHE emission is a pending task. Despite their lower instantaneous sensitivity, EAS arrays with efficient $\gamma$/hadron discrimination can quantify better time-averaged fluxes, integrated over long periods of time, while monitoring for flaring activity, as they continuously drift over large portions of the sky.

The first significant AGN detections at TeV energies were those of Mkn~421 and Mkn~501~\citep{1992Natur.358..477P,1996ApJ...456L..83Q}. These two Markarian galaxies have been extensively studied in the TeV range for over two decades. They are highly variable but remain bright enough over long periods of time to have been detected by EAS arrays like the Tibet Air Shower Array and MILAGRO, both providing first unbiased views of their TeV emission on timescales of years~\citep{2011ApJ...734..110B,2014ApJ...782..110A}. In the last five years, HAWC has been performing an increasingly deeper monitoring of these two BL Lac objects~\citep{2017ApJ...841..100A}. Measurements of their long-term averaged emission are presented in this paper.
 
IACTs have been able to go deeper and beyond these two well-known blazars by implementing sophisticated technologies to achieve lower energy thresholds, around or below 100~GeV, with improved sensitivities. This has permitted to peer through the EBL horizon up to redshifts $z\lesssim 0.9$~\citep{2015ApJ...815L..22A,2015ApJ...815L..23A}, while sampling different types of extragalactic sources, like: 
\begin{itemize}
\item the starbursts galaxies M82 and NGC~253, detected by VERITAS and H.E.S.S. respectively, with fluxes below 1\% of the Crab~\citep{2016CRPhy..17..585O}.
\item four Faranoff-Riley type I {radiogalaxies}:  Centaurus~A (too Southern for HAWC); NGC~1275, the most prominent galaxy of the massive Perseus cluster; 3C~264, newly reported as a VHE source; and M87, the massive central elliptical galaxy in the nearby Virgo cluster. We note the ambiguous classification of IC~310 and PKS~0625--35, referred as RDG in the 3FHL, and as unknown type of AGN in TeVCat~\citep{2008ICRC....3.1341W} and by~\citet{2018Galax...6..116R}.
\item at least sixty {BL Lacertae objects} as VHE $\gamma$-ray sources~\citep{2008ICRC....3.1341W,2017ApJS..232...18A}, mostly high-frequency peaked BL Lac objects (HBL; 50 sources), plus a few intermediate-frequency peaked BLL (IBL; 8 sources, including BL~Lac, W~Comae, and VER~J0521+211, in our sample); and only two Low-frequency peak BL Lacs (LBL). While most of the IACT observations have been reported with sub-TeV thresholds, we note the observations of H~1426+428 by VERITAS and HEGRA at energies between 0.25 and 2.5~TeV, reporting fluxes from 3 to 10\% of the Crab~\citep{2002ApJ...580..104P,2002A&A...384L..23A}.
\item seven {FSRQs}, of which PKS~0736+017 is the only FSRQ detected at VHE with $z<0.3$ and entering the field-of-view of HAWC. This object was found in a flaring state at a flux level of 100~mCrab between 100 and 300~GeV, and showing a steep spectrum~\citep{2017ICRC...35..627C}. 
\end{itemize}

\subsection{Photon Flux Extrapolations\label{extrapol}}
\begin{figure}
\includegraphics[width=\hsize]{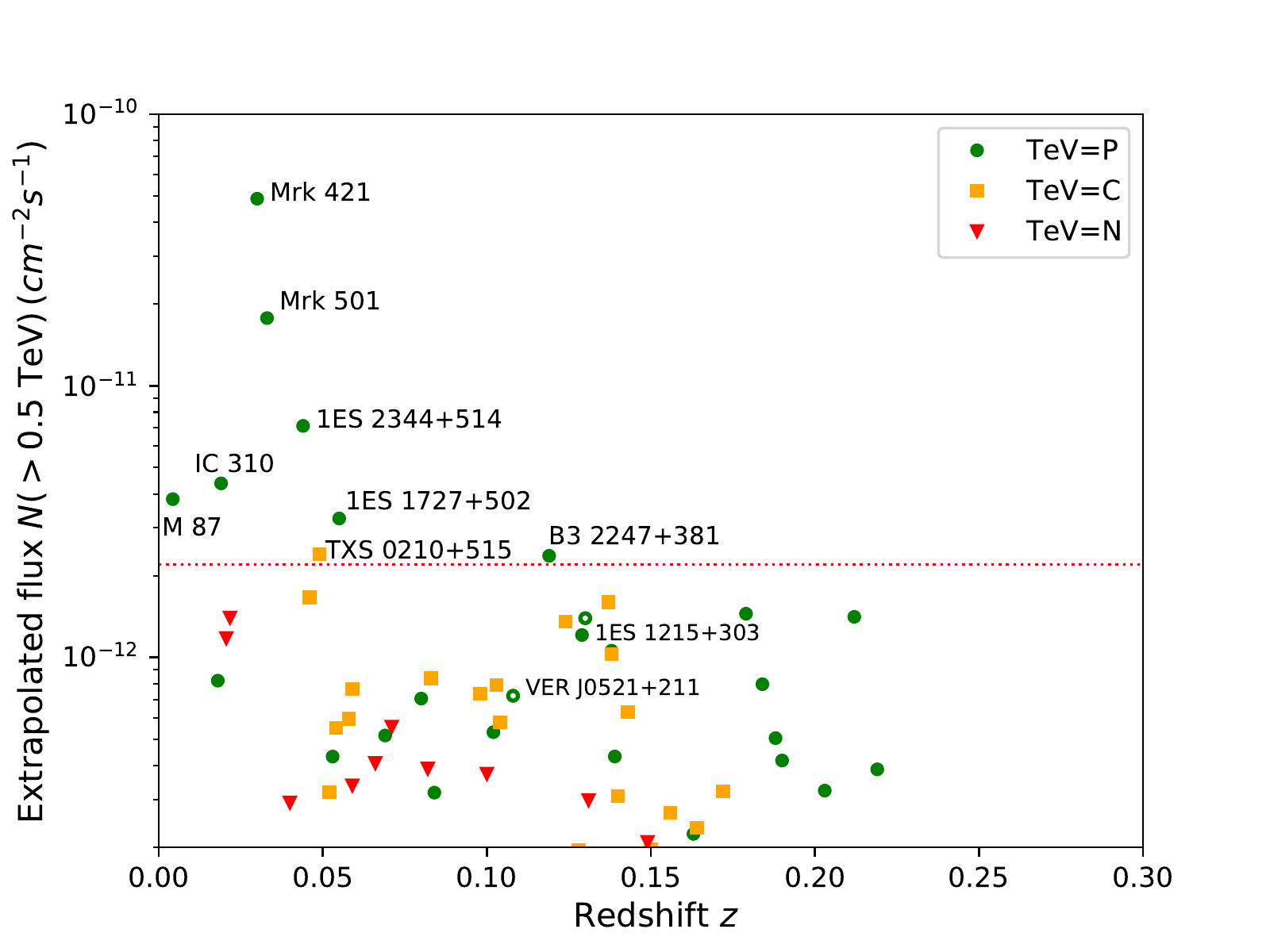}
\caption{Photon flux extrapolations to $E>0.5\,\rm TeV$ using the spectral models of the 3FHL catalog and $\gamma\gamma$ attenuation by the EBL. Colors in the dots indicate the TeV flag assigned in the 3FHL catalog. The horizontal red dotted line represents a flux equivalent 30~mCrab, which we take as a benchmark of potential HAWC detectability. We identify the VHE sources VER~J0521+211 and 1ES~1215+303 with respective green rings below the red line. 
\label{3fhl-xtrapol}}
\end{figure}

{We calculated the expected integral photon flux above 0.5~TeV by extrapolating the spectral models and parameters listed in the 3FHL.} While in HAWC a distinction is made between the {\em intrinsic} spectrum of a source (as emitted) and the related {\em observable} parameters (attenuated by the EBL), {\em Fermi}-LAT spectral models do not need to make this distinction. The difference is minor in most of the LAT energy regime, with only a moderate increase in spectral indices for sources with $z\gtrsim 1$~\citep{2017ApJS..232...18A}. Therefore, we added the attenuation effect of the EBL to the LAT spectral models. The extrapolated fluxes as function of redshift are shown in Figure~\ref{3fhl-xtrapol}. The horizontal red line corresponds to a photon flux of 3\% of the Crab Nebula, indicative of potential HAWC detectability within the current data. 
Mkn~421 and Mkn~501 stand clearly at about an order of magnitude above the red line, with a few other target candidates above the 30~mCrab reference: M87, the nearest radiogalaxy in our sample; IC~310, already reported as a VHE source;  1ES~2344+514, TXS~0210+515, and I~Zw~187 (1ES1727+502), three of the nearest BL Lac objects known; and the mildly distant B3~2247+381, clearly detected by {\em Fermi}-LAT up to 500~GeV. Only TXS~0210+515 is undetected in the VHE range.

We also computed photon flux extrapolations of VHE sources within TeVCat, using simple power laws and approximating the EBL attenuation by a direct exponential in redshift (\S\ref{sec:ebl}). The predictions from this second set are more uncertain, as IACTs observations are by nature short and sparse, reflect different activity states, and spectra are not always fitted above 1~TeV. Five candidates stood out with extrapolated fluxes $N(>0.5\,{\rm TeV})\geq 20~{\rm mCrab}$, four of them in common with the 3FHL extrapolation: M87, Mkn~421, Mkn~501, H~1426+428, and 1ES~2344+514. Results for the sources named in this section are shown and discussed in the following section~\S\ref{sec:3fhl-follow-up}.

\section{HAWC follow-up survey of 3FHL AGN \label{sec:3fhl-follow-up}}
The HAWC follow-up survey consists of a systematic search for TeV $\gamma$-ray emission from each of the 138 AGN selected from the {\em Fermi}-LAT 3FHL catalog. We performed a maximum likelihood test assuming a point source at the location of the presumed 3FHL counterpart with an intrinsic power-law differential flux spectra of index $\alpha$ attenuated by the EBL, 
\begin{equation}
\left(dN\over dE\right)_{\rm obs} = K \left(E/E_0\right)^{-\alpha}\, e^{-\tau(E,z)} \, .  \label{hawc-fit}
\end{equation}
A first analysis run was performed with a fixed spectral index, $\alpha=2.5$, fitting only the normalization $K$ for a pivot energy $E_0=1\,\rm TeV$. From this we obtained:
\begin{itemize}
\item The compilation of $TS$ values and 1~TeV normalizations $K$ for the revised sample of 136 AGN\footnote{3FHL~J1105.8+3944 and 3FHL~J1652.7+4024 excluded.}; 
\item Statistics of significances ($s\equiv \pm \sqrt{TS}$) for source classes and TeV flags. Following Wilk's theorem, under the null hypothesis the behavior of $s$ tends to a Gaussian distribution, of mean $\mu(s)\to 0$ and standard deviation $\sigma(s)\to 1/\sqrt{N}$, for $N$ points;
\item The corresponding list of 2$\sigma$ ($\sim 95\%$ confidence level) upper limits on the normalization, $K_{2\sigma}$, computed under the~\citet{1998PhRvD..57.3873F} approach, that allows only non-negative fluxes;
\item The comparison of flux values and upper limits with the extrapolations performed in \S\ref{extrapol};
\item The characterization of upper limits as function of declination and redshift.
\end{itemize}
For those sources with test-statistic $TS>9$ we then computed optimized spectral fits, allowing both $K$ and $\alpha$ to vary. The discussion of the five sources complying with this criterion is presented in \S\ref{subsec:optimized}.  Finally, we also computed quasi differential limits for all the sources in the sample in three energy intervals, (0.5-2.0)~TeV, (2.0-8.0)~TeV, and (8.0-32)~TeV, following the procedure implemented in~\citet{2017A&A...607A.115I}. The first energy interval, the HAWC overlap with 3FHL, is where we expect AGN emission to be brighter, while the second interval includes the peak of HAWC sensitivity. The (8.0-32.0)~TeV range put bounds where the EBL attenuation is more severe. These quasi-differential limits are computed assuming $\alpha=2.0$ {\em without} consideration of EBL attenuation, as detailed in \S4.3.3 of~\citet{2017A&A...607A.115I}. 

\subsection{The $\alpha=2.5$ AGN search \label{subsec:alpha2.5}}
{The default spectral model for the AGN search was an index $\alpha=2.5$, as used in the 3HWC~\citep{2020arXiv200708582A}. However note that in here the index refers to the intrinsic spectrum and that the observed spectra will be softer. While we know from previous analyses that this index can represent fairly well the data from both Mkn~421 and Mkn~501~\citep{2019ICRC...36..654C}, photon indices in extragalactic 3FHL sources peak in the interval between 2.0 and 2.5~\citep{2017ApJS..232...18A}.}

\subsubsection{Overall results}
\begin{figure}
\includegraphics[width=0.495\hsize]{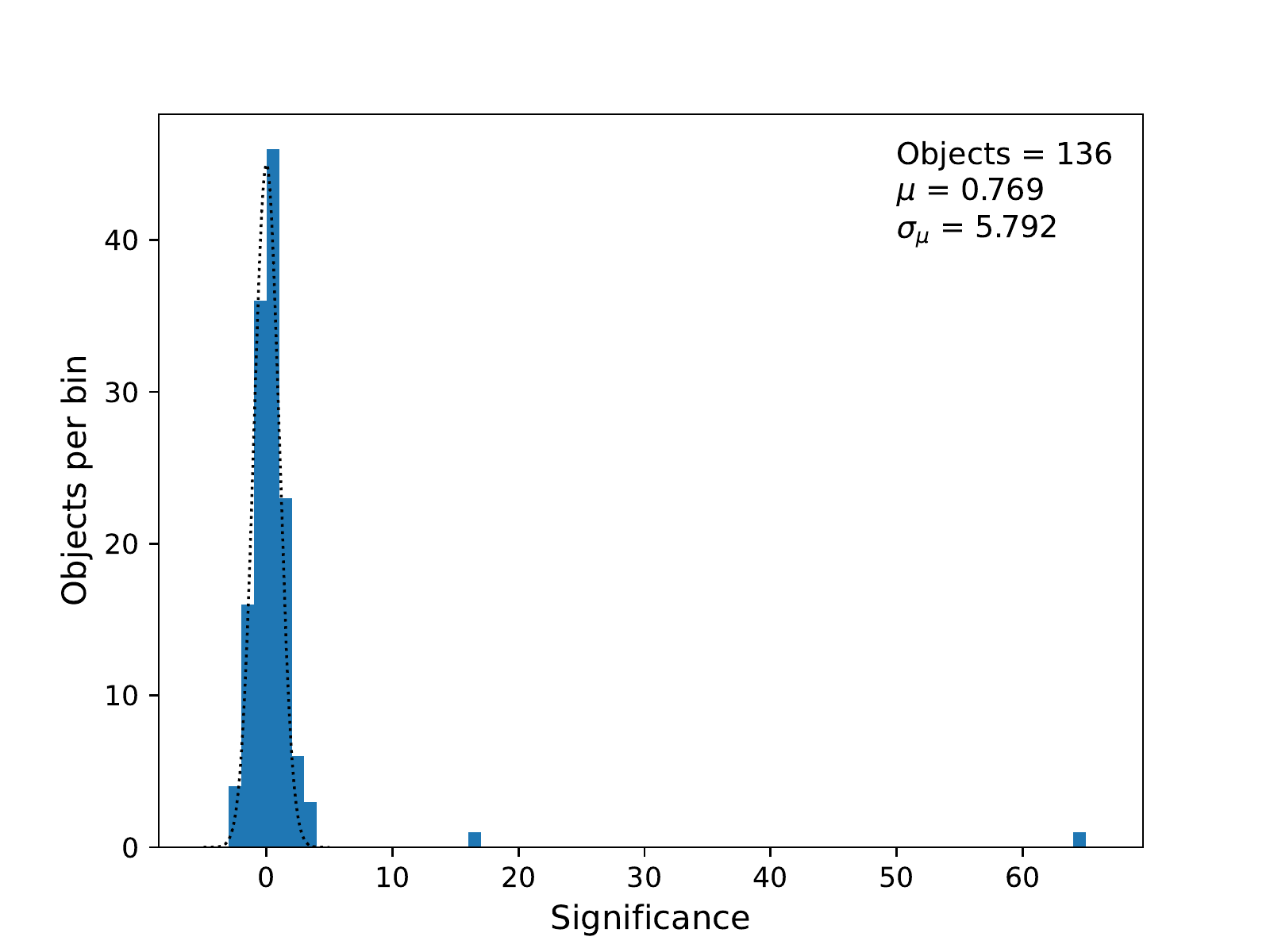} \hfill
\includegraphics[width=0.495\hsize]{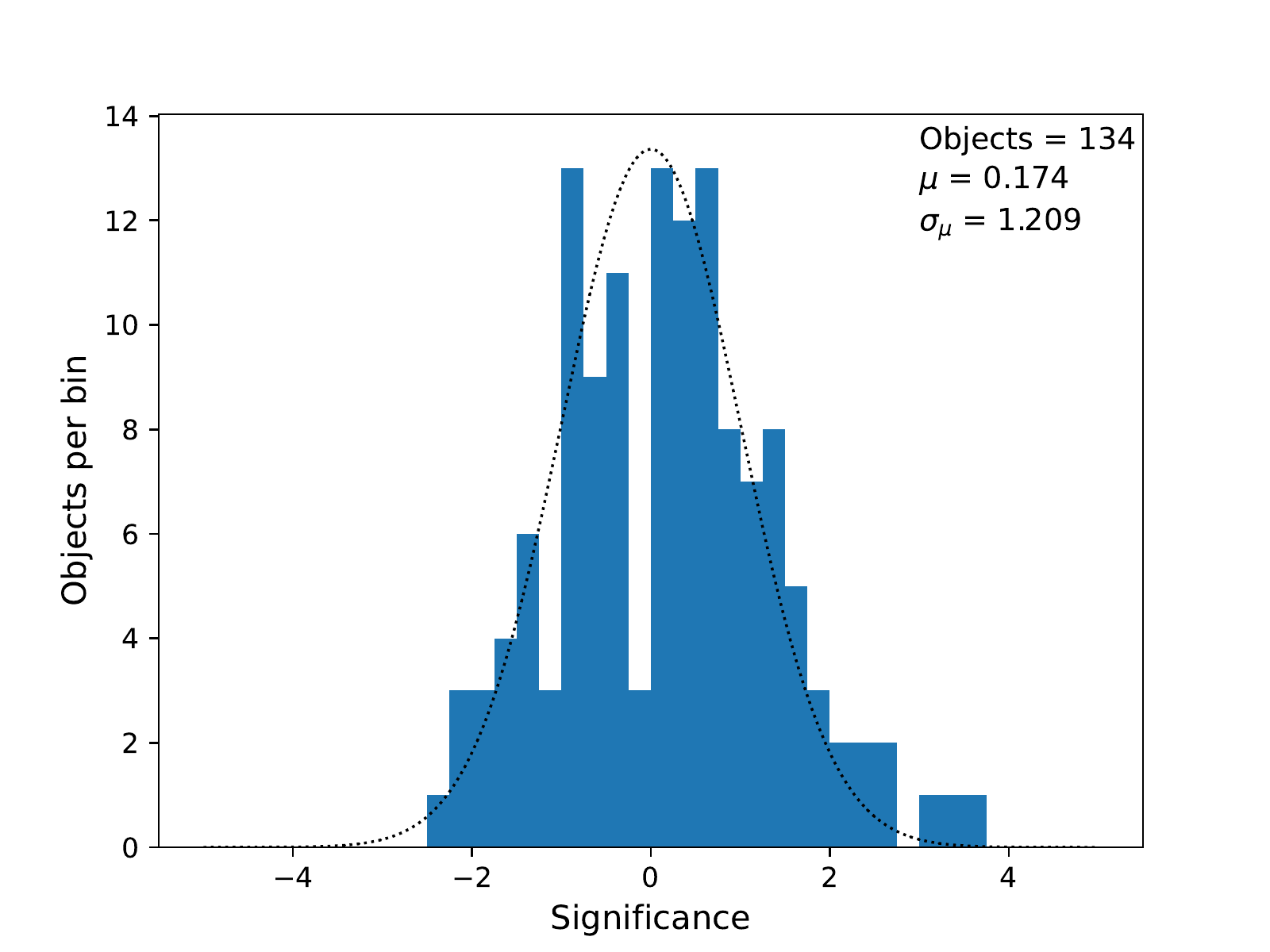}
\caption{Histogram of significances for the sample whole ({\em left}) and for the clean sample, excluding Mkn~421 and Mkn~501 ({\em right}). The point with $\sqrt{TS}=65$ is Mkn~421, while that at 16 is Mkn~501. The three bins at $s>3$ on the right hand-side histogram contain VER~J0521+211, 1ES~1215+303 and M87, each in one bin. 
\label{signif-all}}
\end{figure}

The results for the complete AGN sample are shown in Table~\ref{results-2.5}. The left-hand-side of Figure~\ref{signif-all} shows the histogram of significances for the sample, also displayed as the first entry of Table~\ref{stats-groups}. Mkn~421 is detected with a significance $\sqrt{TS}=+64.6$, and Mkn~501 with $\sqrt{TS}=+16.6$. These two high-significance detections drive up the statistics of the complete sample to an overall $9\sigma$ deviation from the null hypothesis. When removing Mkn~421 and Mkn~501, the joint significance drops to a p-value of 2.2\% (Figure~\ref{signif-all}, right-hand side).

{\startlongtable
\begin{deluxetable*}{lllchrrcc}
\tablecaption{HAWC power-law fits and significances for the revised sample of 136 AGN from the 3FHL catalog. \label{results-2.5}}
\tablehead{\colhead{3FHL source} & \colhead{Counterpart} & \colhead{Class} & \colhead{$z$} & \nocolhead{~$(\alpha_{2000}; \delta_{2000})$} & \colhead{$TS$} & 
\colhead{$\pm\sqrt{TS}$} & \colhead{$K \pm \Delta K$} & \colhead{$K_{2\sigma}$} }
\startdata 
 J0007.9+4711  & RX J0007.9+4711          & bll  & 0.2800 & $  2.000; +47.202$ & $  -4.49$  & $ -2.12$ & $-31.9 \pm 15.1 $ &  9.16 \\
 J0013.8--1855 & RBS 0030                 & bll  & 0.0949 & $  3.484; -18.902$ & $  -0.10$  & $ -0.32$ & $-2.20 \pm  6.81 $ & 11.5 \\
 J0018.6+2946  & RBS 0042                 & bll  & 0.1000 & $  4.616; +29.792$ & $  +0.01$  & $ +0.11$ & $+0.10 \pm  0.93 $ &  1.97 \\
 J0037.8+1239  & NVSS J003750+123818      & bll  & 0.0890 & $  9.462; +12.639$ & $  +0.47$  & $ +0.69$ & $+0.52 \pm  0.76 $ &  2.03 \\
 J0047.9+3947  & B3 0045+395              & bll  & 0.2520 & $ 11.980; +39.816$ & $  -0.18$  & $ -0.42$ & $-2.68 \pm  6.38 $ & 10.3 \\
 J0056.3--0936 & TXS 0053--098            & bll  & 0.1031 & $ 14.084;  -9.608$ & $  -0.21$  & $ -0.46$ & $-1.33 \pm  2.88 $ &  4.53 \\
 J0059.3--0152 & 1RXS J005916.3--015030   & bll  & 0.1440 & $ 14.821;  -1.838$ & $  +0.95$  & $ +0.98$ & $+2.61 \pm  2.67 $ &  7.93 \\
 J0112.1+2245  & S2 0109+22               & BLL  & 0.2650 & $ 18.024; +22.744$ & $  +0.33$  & $ +0.57$ & $+1.99 \pm  3.47 $ &  8.88 \\
 J0123.0+3422  & 1ES 0120+340             & bll  & 0.2720 & $ 20.786; +34.347$ & $  +6.53$  & $ +2.56$ & $+13.1 \pm  5.13 $ & 23.4 \\
 J0131.1+5546  & TXS 0128+554             & bcu  & 0.0365 & $ 22.808; +55.754$ & $  -0.85$  & $ -0.92$ & $-1.61 \pm  1.75 $ &  2.10 \\
 J0152.6+0147  & PMN J0152+0146           & bll  & 0.0800 & $ 28.165;  +1.788$ & $  -0.89$  & $ -0.94$ & $-0.89 \pm  0.94 $ &  1.10 \\
 J0159.5+1047  & RX J0159.5+1047          & bll  & 0.1950 & $ 29.893; +10.785$ & $  +1.09$  & $ +1.04$ & $+2.52 \pm  2.41 $ &  7.31 \\
 J0211.2+1051  & MG1 J021114+1051         & BLL  & 0.2000 & $ 32.805; +10.860$ & $  -0.58$  & $ -0.76$ & $-1.92 \pm  2.51 $ &  3.26 \\
 J0214.5+5145  & TXS 0210+515             & bll  & 0.0490 & $ 33.575; +51.748$ & $  +2.63$  & $ +1.62$ & $+2.43 \pm  1.51 $ &  5.43 \\
 J0216.4+2315  & RBS 0298                 & bll  & 0.2880 & $ 34.134; +23.246$ & $  -0.08$  & $ -0.29$ & $-1.14 \pm  3.97 $ &  6.74 \\
 J0217.1+0836  & ZS 0214+083              & bll  & 0.0850 & $ 34.321;  +8.618$ & $  +4.58$  & $ +2.14$ & $+1.70 \pm  0.80 $ &  3.31 \\
 J0219.1--1723 & 1RXS J021905.8--172503   & bll  & 0.1287 & $ 34.773; -17.420$ & $  +0.01$  & $ +0.10$ & $+1.0 \pm  9.9 $ & 20.9 \\
 J0232.8+2017  & 1ES 0229+200             & bll  & 0.1400 & $ 38.203; +20.288$ & $  -0.28$  & $ -0.53$ & $-0.70 \pm  1.32 $ &  1.98 \\
 J0242.7--0002 & NGC 1068                 & sbg  & 0.0038 & $ 40.670;  -0.013$ & $  +1.45$  & $ +1.21$ & $+0.24 \pm  0.20 $ &  0.65 \\
 J0308.4+0408  & NGC 1218                 & rdg  & 0.0288 & $ 47.109;  +4.111$ & $  +0.24$  & $ +0.49$ & $+0.17 \pm  0.35 $ &  0.87 \\
 J0312.8+3614  & V Zw 326                 & bll  & 0.0710 & $ 48.210; +36.255$ & $  +0.48$  & $ +0.69$ & $+0.53 \pm  0.77 $ &  2.07 \\
 J0316.6+4120  & IC 310                   & RDG  & 0.0189 & $ 49.179; +41.325$ & $  +0.74$  & $ +0.86$ & $+0.31 \pm  0.36 $ &  1.01 \\
 J0319.8+1845  & RBS 0413                 & bll  & 0.1900 & $ 49.966; +18.760$ & $  +0.05$  & $ +0.22$ & $+0.45 \pm  2.08 $ &  4.64 \\
 J0319.8+4130  & NGC 1275                 & RDG  & 0.0176 & $ 49.951; +41.512$ & $  +0.36$  & $ +0.60$ & $+0.21 \pm  0.35 $ &  0.91 \\
 J0325.6--1646 & RBS 0421                 & bll  & 0.2910 & $ 51.421; -16.771$ & $  +0.07$  & $ +0.27$ & $+12.6 \pm 46.4 $ & 107 \\
 J0326.3+0226  & 1H 0323+022              & bll  & 0.1470 & $ 51.558;  +2.421$ & $  -0.22$  & $ -0.47$ & $-1.02 \pm  2.16 $ &  3.35 \\
 J0334.3+3920  & 4C+39.12                 & rdg  & 0.0203 & $ 53.577; +39.357$ & $  -0.91$  & $ -0.96$ & $-0.33 \pm  0.34 $ &  0.40 \\
 J0336.4--0348 & 1RXS J033623.3--034727   & bll  & 0.1618 & $ 54.099;  -3.794$ & $  +0.19$  & $ +0.44$ & $+1.64 \pm  3.75 $ &  9.10 \\
 J0339.2--1736 & PKS 0336--177            & bcu  & 0.0656 & $ 54.807; -17.600$ & $  +0.27$  & $ +0.52$ & $+1.83 \pm  3.55 $ &  8.97 \\
 J0349.3--1159 & 1ES 0347--121            & bll  & 0.1850 & $ 57.347; -11.991$ & $  +0.74$  & $ +0.86$ & $+9.00 \pm 10.5 $ & 30.0 \\
 J0416.8+0105  & 1ES 0414+009             & bll  & 0.2870 & $ 64.219;  +1.090$ & $  +1.71$  & $ +1.31$ & $+9.16 \pm  7.00 $ & 22.9 \\
 J0424.7+0036  & PKS 0422+00              & bll  & 0.2680 & $ 66.195;  +0.602$ & $  +1.60$  & $ +1.27$ & $+8.17 \pm  6.46 $ & 21.1 \\
 J0521.7+2112  & TXS 0518+211   & bll  & 0.1080 & $ 80.442; +21.214$ & $  +9.49$  & $ +3.08$ & $+2.85 \pm  0.93 $ &  4.72 \\
 J0602.0+5316  & GB6 J0601+5315           & bcu  & 0.0520 & $ 90.502; +53.267$ & $  -0.17$  & $ -0.42$ & $-0.79 \pm  1.88 $ &  3.01 \\
 J0617.6--1715 & TXS 0615--172            & bll  & 0.0980 & $ 94.389; -17.257$ & $  +1.51$  & $ +1.23$ & $+7.46 \pm  6.07 $ & 19.7 \\
 J0648.7+1517  & RX J0648.7+1516          & bll  & 0.1790 & $102.198; +15.274$ & $  +0.26$  & $ +0.51$ & $+1.00 \pm  1.97 $ &  4.92 \\
 J0650.7+2503  & 1ES 0647+250             & bll  & 0.2030 & $102.694; +25.050$ & $  +0.01$  & $ +0.10$ & $+0.24 \pm  2.43 $ &  5.12 \\
 J0656.2+4235  & 4C +42.22                & bll  & 0.0590 & $104.044; +42.617$ & $  -0.65$  & $ -0.81$ & $-0.71 \pm  0.88 $ &  1.13 \\
 J0725.8--0056 & PKS 0723--008            & bcu  & 0.1270 & $111.461;  -0.916$ & $  -2.76$  & $ -1.66$ & $-3.48 \pm  2.09 $ &  1.58 \\
 J0730.4+3307  & 1RXS J073026.0+330727    & bll  & 0.1120 & $112.609; +33.123$ & $  +0.96$  & $ +0.98$ & $+1.21 \pm  1.23 $ &  3.66 \\
 J0739.3+0137  & PKS 0736+01              & fsrq & 0.1910 & $114.825;  +1.618$ & $  -0.96$  & $ -0.98$ & $-3.40 \pm  3.47 $ &  3.93 \\
 J0753.1+5354  & 4C +54.15                & bll  & 0.2000 & $118.256; +53.883$ & $  +0.78$  & $ +0.88$ & $+15.2 \pm 17.3 $ & 49.6 \\
 J0757.1+0957  & PKS 0754+100             & bll  & 0.2660 & $119.278;  +9.943$ & $  -0.12$  & $ -0.34$ & $-1.39 \pm  4.04 $ &  6.80 \\
 J0809.7+3457  & B2 0806+35               & bll  & 0.0830 & $122.412; +34.927$ & $  +0.09$  & $ +0.30$ & $+0.26 \pm  0.87 $ &  2.03 \\
 J0809.8+5218  & 1ES 0806+524             & bll  & 0.1371 & $122.455; +52.316$ & $  +0.27$  & $ +0.52$ & $+3.59 \pm  6.9 $ & 17.3 \\
 J0816.4--1311 & PMN J0816--1311          & bll  & 0.0460 & $124.113; -13.198$ & $  -4.67$  & $ -2.16$ & $-3.23 \pm  1.49 $ &  0.89 \\
 J0816.4+5739  & SBS 0812+578             & bll  & 0.2940 & $124.095; +57.653$ & $  -0.55$  & $ -0.74$ & $-2.32 \pm  3.14 $ &  4.16 \\
 J0816.9+2050  & SDSS J081649.78+205106.4 & bll  & 0.0583 & $124.207; +20.852$ & $  -2.72$  & $ -1.65$ & $-0.77 \pm  0.47 $ &  0.36 \\
 J0828.3+4153  & GB6 B0824+4203           & bll  & 0.2262 & $127.059; +41.898$ & $  +0.84$  & $ +0.92$ & $+5.80 \pm  6.33 $ & 18.4 \\
 J0831.8+0429  & PKS 0829+046             & bll  & 0.1738 & $127.954;  +4.494$ & $  +2.00$  & $ +1.41$ & $+3.62 \pm  2.56 $ &  8.72 \\
 J0847.2+1134  & RX J0847.1+1133          & bll  & 0.1982 & $131.804; +11.564$ & $  +0.09$  & $ +0.29$ & $+0.72 \pm  2.45 $ &  5.65 \\
 J0850.6+3454  & RX J0850.5+3455          & bll  & 0.1450 & $132.651; +34.923$ & $  +1.43$  & $ +1.20$ & $+2.33 \pm  1.95 $ &  6.20 \\
 J0908.9+2311  & RX J0908.9+2311          & bll  & 0.2230 & $137.253; +23.187$ & $  -2.15$  & $ -1.47$ & $-3.98 \pm  2.71 $ &  2.32 \\
 J0912.4+1555  & SDSS J091230.61+155528.0 & bll  & 0.2120 & $138.128; +15.924$ & $  -0.47$  & $ -0.69$ & $-1.73 \pm  2.52 $ &  3.54 \\
 J0930.4+4952  & 1ES 0927+500             & bll  & 0.1867 & $142.657; +49.840$ & $  -4.58$  & $ -2.14$ & $-20.4 \pm  9.53 $ &  5.60 \\
 J1015.0+4926  & 1H 1013+498              & bll  & 0.2120 & $153.767; +49.434$ & $  +0.03$  & $ +0.19$ & $+2.14 \pm 11.5 $ & 25.1 \\
 J1027.0--1749 & 1RXS J102658.5--174905   & bll  & 0.2670 & $156.744; -17.816$ & $  +0.10$  & $ +0.32$ & $+14.6 \pm 45.5 $ & 105 \\
 J1041.7+3900  & B3 1038+392              & bll  & 0.2084 & $160.455; +39.022$ & $  -0.51$  & $ -0.71$ & $-3.15 \pm  4.43 $ &  6.01 \\
 J1053.6+4930  & GB6 J1053+4930           & bll  & 0.1404 & $163.434; +49.499$ & $  +0.31$  & $ +0.56$ & $+2.97 \pm  5.33 $ & 13.8 \\
 J1058.6+5628  & TXS 1055+567             & BLL  & 0.1433 & $164.657; +56.470$ & $  -1.62$  & $ -1.27$ & $-15.5 \pm 12.2 $ & 11.7 \\
 J1100.3+4020  & RX J1100.3+4019          & bll  & 0.2250 & $165.088; +40.324$ & $  +2.67$  & $ +1.63$ & $+9.08 \pm  5.56 $ & 20.3 \\
 J1104.4+3812 & Mkn 421                  & BLL  & 0.0310 & $166.114; +38.209$ & $+4166.97$  & $+64.55$ & $+29.5 \pm  0.5 $ & 30.6 \\
 J1117.0+2014  & RBS 0958                 & bll  & 0.1380 & $169.276; +20.235$ & $  +5.28$  & $ +2.30$ & $+3.00 \pm  1.31 $ &  5.63 \\
 J1120.8+4212  & RBS 0970                 & bll  & 0.1240 & $170.200; +42.203$ & $  +1.42$  & $ +1.19$ & $+2.73 \pm  2.29 $ &  7.29 \\
 J1125.9--0743 & 1RXS J112551.6--074219   & bll  & 0.2790 & $171.467;  -7.706$ & $  -2.74$  & $ -1.65$ & $-23.2 \pm 14.0 $ & 10.8 \\
 J1136.8+2549  & RX J1136.8+2551          & bll  & 0.1560 & $174.209; +25.848$ & $  -1.13$  & $ -1.06$ & $-1.74 \pm  1.64 $ &  1.75 \\
 J1140.5+1528  & NVSS J114023+152808      & bll  & 0.2443 & $175.098; +15.469$ & $  -0.59$  & $ -0.77$ & $-2.41 \pm  3.13 $ &  4.00 \\
 J1142.0+1546  & MG1 J114208+1547         & bll  & 0.2990 & $175.532; +15.798$ & $  -0.93$  & $ -0.96$ & $-4.08 \pm  4.24 $ &  4.94 \\
 J1145.0+1935  & 3C~264                   & rdg  & 0.0216 & $176.271; +19.606$ & $  +3.61$  & $ +1.90$ & $+0.44 \pm  0.24 $ &  0.92 \\
 J1150.3+2418  & OM 280                   & bll  & 0.2000 & $177.580; +24.298$ & $  +7.15$  & $ +2.67$ & $+6.24 \pm  2.33 $ & 10.9 \\
 J1154.1--0010 & 1RXS J115404.9--001008   & bll  & 0.2535 & $178.519;  -0.169$ & $  -0.28$  & $ -0.53$ & $-3.27 \pm  6.19 $ &  9.46 \\
 J1204.2--0709 & 1RXS J120417.0--070959   & bll  & 0.1850 & $181.069;  -7.169$ & $  -2.30$  & $ -1.52$ & $-9.58 \pm  6.31 $ &  5.09 \\
 J1217.9+3006  & 1ES 1215+303             & bll  & 0.1300 & $184.467; +30.117$ & $ +11.36$  & $ +3.37$ & $+4.64 \pm  1.38 $ &  7.39 \\
 J1219.7--0312 & 1RXS J121946.0--031419   & bll  & 0.2988 & $184.940;  -3.240$ & $  -0.94$  & $ -0.97$ & $-10.0 \pm 10.3 $ & 12.4 \\
 J1221.3+3010  & PG 1218+304              & bll  & 0.1837 & $185.341; +30.177$ & $  +5.02$  & $ +2.24$ & $+5.23 \pm  2.34 $ &  9.92 \\
 J1221.5+2813  & W Comae                  & bll  & 0.1029 & $185.382; +28.233$ & $  +6.03$  & $ +2.45$ & $+2.33 \pm  0.95 $ &  4.24 \\
 J1224.4+2436  & MS 1221.8+2452           & bll  & 0.2187 & $186.101; +24.607$ & $  +0.55$  & $ +0.74$ & $+1.99 \pm  2.68 $ &  7.32 \\
 J1229.2+0201  & 3C 273                   & FSRQ & 0.1583 & $187.278;  +2.052$ & $  -1.97$  & $ -1.40$ & $-3.48 \pm  2.48 $ &  2.15 \\
 J1230.2+2517  & ON~246                   & bll  & 0.1350 & $187.559; +25.302$ & $  +0.43$  & $ +0.66$ & $+0.86 \pm  1.31 $ &  3.46 \\
 J1230.8+1223  & M87                      & rdg  & 0.0042 & $187.706; +12.391$ & $ +12.93$  & $ +3.60$ & $+0.56 \pm  0.16 $ &  0.88 \\
 J1231.4+1422  & GB6 J1231+1421           & bll  & 0.2559 & $187.850; +14.357$ & $  -0.09$  & $ -0.30$ & $-1.01 \pm  3.40 $ &  5.82 \\
 J1231.7+2847  & B2 1229+29               & bll  & 0.2360 & $187.932; +28.797$ & $  +0.00$  & $ +0.05$ & $+0.18 \pm  3.30 $ &  6.75 \\
 J1253.7+0328  & MG1 J125348+0326         & bll  & 0.0657 & $193.446;  +3.442$ & $  -3.80$  & $ -1.95$ & $-1.36 \pm  0.70 $ &  0.46 \\
 J1256.2--1146 & PMN J1256--1146          & bcu  & 0.0579 & $194.066; -11.777$ & $  +0.26$  & $ +0.51$ & $+0.84 \pm  1.63 $ &  4.11 \\
 J1310.3--1158 & TXS 1307--117            & bll  & 0.1400 & $197.552; -11.963$ & $  -3.50$  & $ -1.87$ & $-11.4 \pm  6.10 $ &  4.11 \\
 J1341.2+3959  & RBS 1302                 & bll  & 0.1715 & $205.271; +39.996$ & $  -0.97$  & $ -0.99$ & $-3.35 \pm  3.40 $ &  3.92 \\
 J1402.6+1559  & MC 1400+162              & bll  & 0.2440 & $210.685; +15.999$ & $  +1.78$  & $ +1.33$ & $+4.14 \pm  3.10 $ & 10.3 \\
 J1411.8+5249  & SBS 1410+530             & bcu  & 0.0765 & $212.956; +52.817$ & $  +0.24$  & $ +0.49$ & $+1.40 \pm  2.84 $ &  7.12 \\
 J1418.0+2543  & 1E 1415.6+2557           & bll  & 0.2363 & $214.486; +25.724$ & $  +0.27$  & $ +0.52$ & $+1.60 \pm  3.07 $ &  7.78 \\
 J1419.4+0444  & SDSS J141927.49+044513.7 & bll  & 0.1430 & $214.865;  +4.754$ & $  +0.16$  & $ +0.40$ & $+0.74 \pm  1.85 $ &  4.50 \\
 J1419.7+5423  & OQ 530                   & bll  & 0.1525 & $214.944; +54.387$ & $  +0.30$  & $ +0.55$ & $+5.94 \pm 10.8 $ & 27.6 \\
 J1428.5+4240  & H 1426+428               & bll  & 0.1292 & $217.136; +42.673$ & $  +1.58$  & $ +1.26$ & $+3.18 \pm  2.54 $ &  8.18 \\
 J1436.9+5639  & RBS 1409                 & bll  & 0.1500 & $219.240; +56.657$ & $  +2.38$  & $ +1.54$ & $+21.0 \pm 13.6 $ & 48.1 \\
 J1442.8+1200  & 1ES 1440+122             & bll  & 0.1631 & $220.701; +12.011$ & $  +3.53$  & $ +1.88$ & $+3.37 \pm  1.80 $ &  6.95 \\
 J1449.5+2745  & B2.2 1447+27             & bll  & 0.2272 & $222.386; +27.773$ & $  +0.24$  & $ +0.49$ & $+1.49 \pm  3.03 $ &  7.46 \\
 J1500.9+2238  & MS 1458.8+2249           & bll  & 0.2350 & $225.258; +22.635$ & $  +3.06$  & $ +1.75$ & $+5.10 \pm  2.91 $ & 11.0 \\
 J1508.7+2708  & RBS 1467                 & bll  & 0.2700 & $227.178; +27.152$ & $  +0.15$  & $ +0.38$ & $+1.48 \pm  3.89 $ &  9.29 \\
 J1512.2+0203  & PKS 1509+022             & fsrq & 0.2195 & $228.066;  +2.055$ & $  -1.84$  & $ -1.36$ & $-5.73 \pm  4.23 $ &  3.81 \\
 J1518.5+4044  & GB6 J1518+4045           & bll  & 0.0652 & $229.662; +40.750$ & $  +0.91$  & $ +0.96$ & $+0.83 \pm  0.87 $ &  2.57 \\
 J1531.9+3016  & RX J1531.9+3016          & bll  & 0.0653 & $233.009; +30.275$ & $  +0.03$  & $ +0.18$ & $+0.10 \pm  0.57 $ &  1.25 \\
 J1543.6+0452  & CGCG 050--083            & bcu  & 0.0400 & $235.891;  +4.872$ & $  -5.82$  & $ -2.41$ & $-1.04 \pm  0.43 $ &  0.24 \\
 J1554.2+2010  & 1ES 1552+203             & bll  & 0.2223 & $238.601; +20.190$ & $  +1.76$  & $ +1.33$ & $+3.50 \pm  2.64 $ &  8.81 \\
 J1603.8+1103  & MG1 J160340+1106         & bll  & 0.1430 & $240.925; +11.097$ & $  -0.17$  & $ -0.41$ & $-0.61 \pm  1.50 $ &  2.41 \\
 J1615.4+4711  & TXS 1614+473             & fsrq & 0.1987 & $243.922; +47.187$ & $  -1.08$  & $ -1.04$ & $-8.29 \pm  7.99 $ &  8.82 \\
 J1643.5--0646 & NVSS J164328--064619     & bcu  & 0.0820 & $250.871;  -6.772$ & $  -0.00$  & $ -0.01$ & $-0.02 \pm  1.61 $ &  3.23 \\
 J1647.6+4950  & SBS 1646+499             & bll  & 0.0475 & $251.895; +49.833$ & $  +0.10$  & $ +0.32$ & $+0.38 \pm  1.21 $ &  2.82 \\
 J1653.8+3945  & Mkn 501                  & BLL  & 0.0330 & $253.468; +39.760$ & $+276.97$  & $+16.64$ & $+7.74 \pm  0.49 $ &  8.72 \\
 J1719.2+1745  & PKS 1717+17              & bll  & 0.1370 & $259.804; +17.752$ & $  -0.32$  & $ -0.56$ & $-0.72 \pm  1.28 $ &  1.90 \\
 J1725.4+5851  & 7C 1724+5854             & bll  & 0.2970 & $261.396; +58.861$ & $  -1.94$  & $ -1.39$ & $-99.0 \pm 71.0 $ & 63.5 \\
 J1728.3+5013  & I Zw 187                 & bll  & 0.0550 & $262.078; +50.220$ & $  -0.05$  & $ -0.22$ & $-0.32 \pm  1.44 $ &  2.57 \\
 J1730.8+3715  & GB6 J1730+3714           & bll  & 0.2040 & $262.696; +37.249$ & $  -0.21$  & $ -0.45$ & $-1.71 \pm  3.78 $ &  5.86 \\
 J1744.0+1935  & S3 1741+19               & bll  & 0.0830 & $265.991; +19.586$ & $  +1.62$  & $ +1.27$ & $+0.83 \pm  0.66 $ &  2.17 \\
 J1745.6+3950  & B2 1743+39C              & bll  & 0.2670 & $266.407; +39.859$ & $  +0.04$  & $ +0.21$ & $+1.47 \pm  7.11 $ & 15.5 \\
 J1813.5+3144  & B2 1811+31               & bll  & 0.1170 & $273.397; +31.738$ & $  -1.68$  & $ -1.30$ & $-1.59 \pm  1.23 $ &  1.14 \\
 J1917.7--1921 & 1H 1914--194             & bll  & 0.1370 & $289.437; -19.359$ & $  +0.01$  & $ +0.12$ & $+1.64 \pm 13.6 $ & 28.9 \\
 J2000.4--1327 & NVSS J200042--132532     & fsrq & 0.2220 & $300.176; -13.426$ & $  -0.76$  & $ -0.87$ & $-15.2 \pm 17.4 $ & 21.1 \\
 J2014.4--0047 & PMN J2014--0047          & bll  & 0.2310 & $303.619;  -0.790$ & $  +1.23$  & $ +1.11$ & $+6.12 \pm  5.51 $ & 17.1 \\
 J2039.4+5219  & 1ES 2037+521             & bll  & 0.0540 & $309.848; +52.331$ & $  -0.94$  & $ -0.97$ & $-1.70 \pm  1.75 $ &  2.04 \\
 J2042.0+2428  & MG2 J204208+2426         & bll  & 0.1040 & $310.525; +24.448$ & $  +0.58$  & $ +0.76$ & $+0.67 \pm  0.89 $ &  2.44 \\
 J2055.0+0014  & RGB J2054+002            & bll  & 0.1508 & $313.737;  +0.261$ & $  -0.26$  & $ -0.51$ & $-1.27 \pm  2.51 $ &  3.78 \\
 J2108.8--0251 & TXS 2106--030            & bll  & 0.1490 & $317.187;  -2.843$ & $  -0.55$  & $ -0.74$ & $-2.23 \pm  3.01 $ &  4.03 \\
 J2143.5+1742  & OX 169                   & fsrq & 0.2110 & $325.898; +17.730$ & $  +0.18$  & $ +0.42$ & $+1.03 \pm  2.43 $ &  5.82 \\
 J2145.8+0718  & MS 2143.4+0704           & bll  & 0.2350 & $326.468;  +7.324$ & $  -1.02$  & $ -1.01$ & $-3.65 \pm  3.61 $ &  4.14 \\
 J2150.2--1412 & TXS 2147--144            & bll  & 0.2290 & $327.565; -14.181$ & $  +0.01$  & $ +0.10$ & $+2.02 \pm 20.2 $ & 42.7 \\
 J2202.7+4216  & BL Lacertae              & BLL  & 0.0690 & $330.680; +42.278$ & $  -0.25$  & $ -0.50$ & $-0.50 \pm  1.0 $ &  1.53 \\
 J2250.0+3825  & B3 2247+381              & bll  & 0.1190 & $342.524; +38.410$ & $  -0.27$  & $ -0.52$ & $-0.86 \pm  1.67 $ &  2.54 \\
 J2252.0+4031  & MITG J2252+4030          & bll  & 0.2290 & $343.004; +40.508$ & $  -0.78$  & $ -0.88$ & $-5.03 \pm  5.71 $ &  6.92 \\
 J2314.0+1445  & RGB J2313+147            & bll  & 0.1625 & $348.489; +14.740$ & $  +1.37$  & $ +1.17$ & $+1.97 \pm  1.68 $ &  5.29 \\
 J2322.6+3436  & TXS 2320+343             & bll  & 0.0980 & $350.683; +34.604$ & $  +0.01$  & $ +0.09$ & $+0.09 \pm  1.06 $ &  2.23 \\
 J2323.8+4210  & 1ES 2321+419             & bll  & 0.0590 & $350.967; +42.183$ & $  -3.79$  & $ -1.95$ & $-1.64 \pm  0.84 $ &  0.57 \\
 J2329.2+3755  & NVSS J232914+375414      & bll  & 0.2640 & $352.309; +37.904$ & $  +0.04$  & $ +0.19$ & $+1.17 \pm  6.03 $ & 13.3 \\
 J2338.9+2123  & RX J2338.8+2124          & bll  & 0.2910 & $354.735; +21.411$ & $  +2.91$  & $ +1.71$ & $+6.76 \pm  3.97 $ & 14.7 \\
 J2346.6+0705  & TXS 2344+068             & bll  & 0.1720 & $356.666;  +7.085$ & $  +2.24$  & $ +1.50$ & $+3.33 \pm  2.23 $ &  7.82 \\
 J2347.0+5142  & 1ES 2344+514             & bll  & 0.0440 & $356.770; +51.705$ & $  +2.09$  & $ +1.45$ & $+1.93 \pm  1.34 $ &  4.62 \\
 J2356.2+4035  & NVSS J235612+403648      & bll  & 0.1310 & $359.053; +40.613$ & $  +0.01$  & $ +0.10$ & $+0.22 \pm  2.22 $ &  4.64 \\
 J2359.3--2049 & TXS 2356--210            & bll  & 0.0960 & $359.831; -20.799$ & $  -0.06$  & $ -0.23$ & $-1.97 \pm  8.42 $ & 15.0 \\

\enddata
\tablecomments{
Test-statistics ($TS$) and significances ($\pm\sqrt{TS}$) are estimated allowing fluxes to be positive or negative.
$K$ is the power-law normalization a 1~TeV in units of $10^{-12}\,\rm TeV^{-1}\, cm^{-2}\, s^{-1}$, and $K_{2\sigma}$ its corresponding 2$\sigma$ upper limit in the same units, computed using the method of \citet{1998PhRvD..57.3873F}.} 
\end{deluxetable*}}

{Three more objects showed test-statistic $TS>9$: M87 ($TS=12.9$); 1ES~1215+303 ($TS=11.4$); and VER~J0521+211 ($TS=9.5$).  We note that the random probability of having three $TS>9$ values in 134 trials, once the two Markarians are excluded, is $8.5\times 10^{-4}$.  These three fainter sources have favorable declinations for HAWC, $\left |\delta -19^\circ \right | \lesssim 10^\circ$, allowing for transits of over 6~hours. Spectral fits for the five $TS>9$ sources are presented in~\S\ref{subsec:optimized}}.

\subsubsection{Upper limits and sensitivity}

\begin{figure}
\includegraphics[width=\hsize]{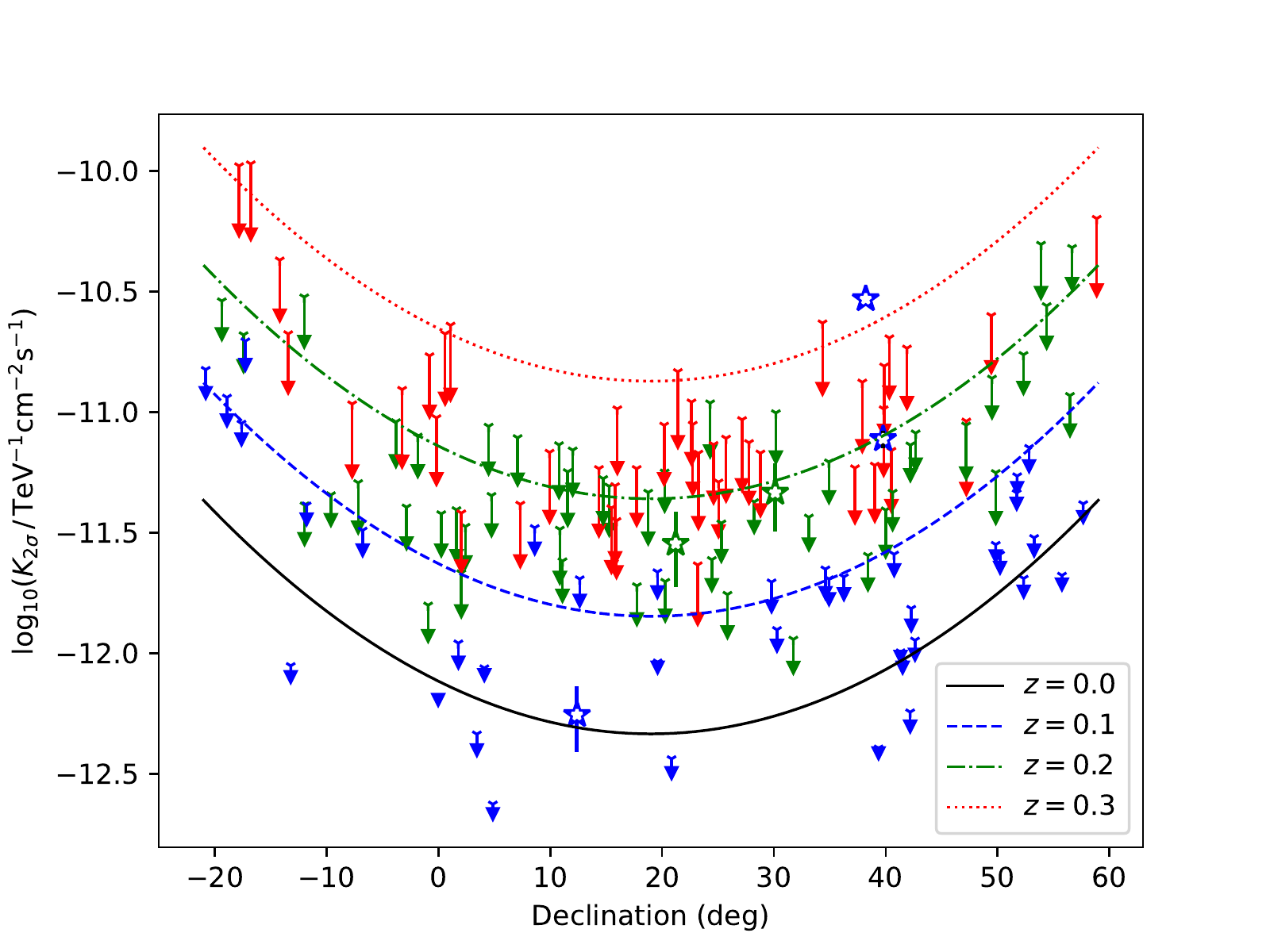}
\caption{Fit of $K_{2\sigma}$, the upper-limit normalizations at 1~TeV, as function of declination and redshift for the ``clean'' sample. The curved dashed lines are the fits given by eq.~\ref{sensitivity} to the upper-limits at the redshifts indicated in the plot. Blue arrows are upper limits for AGN with $z\leq 0.1$, green for $0.1<z\leq 0.2$ and red for $0.2<z\leq 0.3$. Measured values of $K$ for the $TS>9$ sources are indicated by stars. They are at declinations $12^\circ$ (M87, in blue), $21^\circ$ (VER~J0521+211, in green), $30^\circ$ (1ES~1215+303, in green), and close to $40^\circ$ (Mkn~421 as the blue star above the red curve and Mkn~501 the blue star intersecting the green curve). Statistical errors bars are shown, thought smaller than the markers for the Markarian sources. 
\label{ulims-fit}}
\end{figure}

The $K_{2\sigma}$ upper limits span a large range of values: from $2.4\times 10^{-13}\,\rm TeV^{-1}cm^{-2}s^{-1}$, (3FHL~J1543.6+0452 at $z=0.040$), to $10^{-10}\,\rm TeV^{-1}cm^{-2}s^{-1}$ (3FHL~J0325.6--1646 at $z=0.291$). Upper limits are sensitive to both the declination and redshift of the source, as shown in Figure~\ref{ulims-fit}. {We performed a fit to the set of $K_{2\sigma}$ values that assumes a Gaussian dependence in declination and exponential in redshift. Three parameters were computed: the normalization limit at the reference point $\delta=19^\circ$ and $z=0$, the angular width in declination, and the redshift exponential scale. The fit obtained is,}
\begin{equation}
\log_{10}K_{2\sigma} = -12.33 + {1\over 2\ln 10} \left(\delta-19^\circ \over 18.93^{\circ} \right)^{2}+ {1\over \ln 10} \left(z\over 0.089 \right)  \, . \label{sensitivity} 
\end{equation}
{This fit excludes the five $TS>9$ sources. Its correlation coefficient is $r=+0.870$ and the dispersion with the data 0.241~dex. The value for the fit of $K_{2\sigma}$ at $z=0,\,\delta =19^\circ$ corresponds to 14~mCrab, a factor of two lower than the predefined depth of the survey. This would apply to near and optimally located sources only. The sensitivity of the survey degrades with a Gaussian angle of $18.9^\circ$, leading to a 11\% response at $40^\circ$ from zenith. The dependence of the upper limit normalizations with redshift gives $z_h=0.089$, matching that of an observed power-law spectrum integrated from $E_0 = 0.63 \,\rm TeV$, close to the initial assumption $E_0 = 0.5 \,\rm TeV$ (\S\ref{sec:ebl}). }

\subsubsection{Comparison with LAT extrapolations \label{sec:extra-compa}}
\begin{figure}
\includegraphics[width=\hsize]{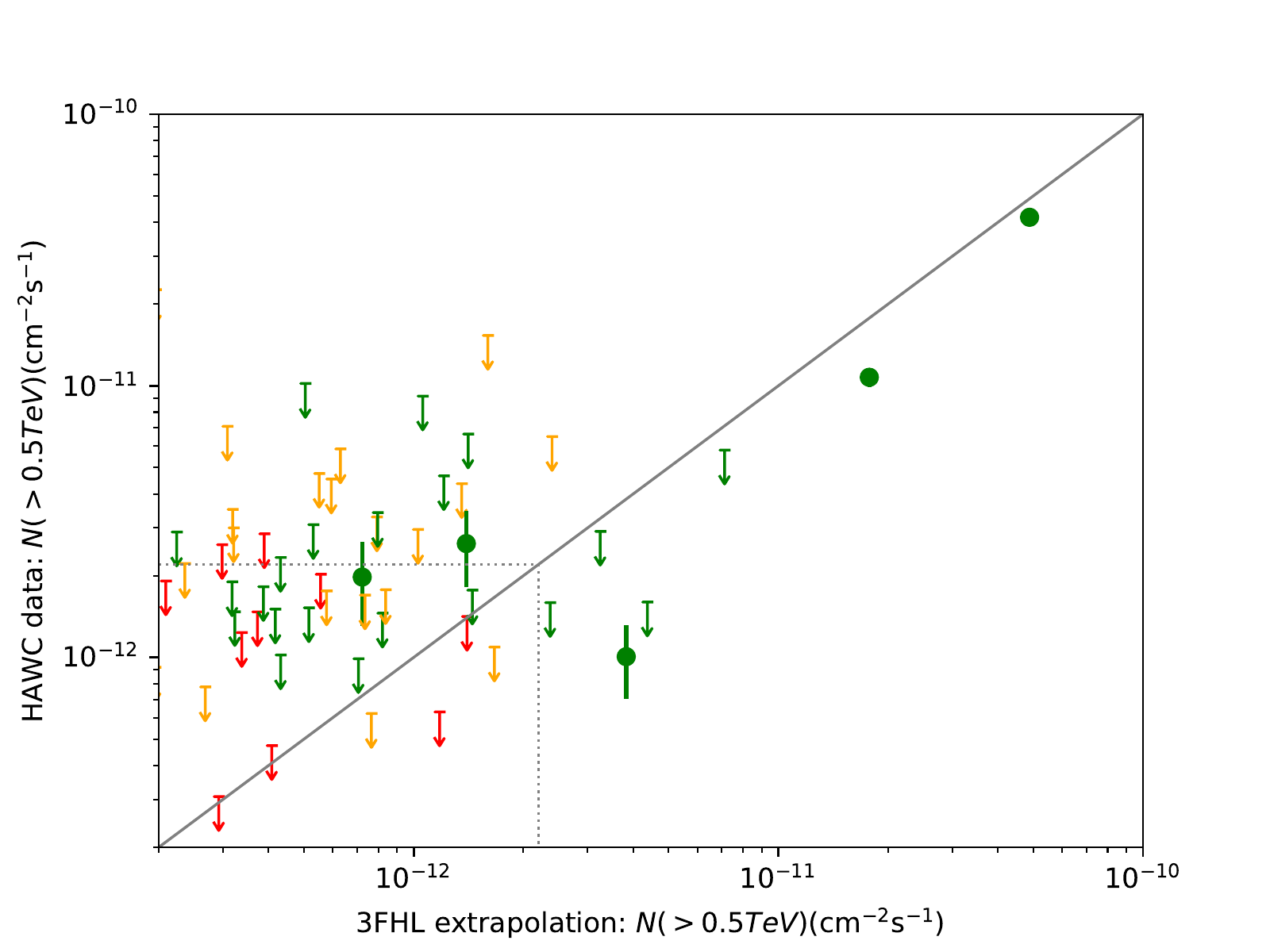}
\caption{Flux extrapolations of 3FHL spectra versus HAWC measurements. Color code indicates the TeV flag: green TeV=P; orange TeV=C; red TeV=N. The green points are the measured fluxes of the five objects with $TS>9$, the rest of the sample indicated by upper limits. Errors shown are statistical only.
\label{extrapol-ulims}} 
\end{figure}

In Figure~\ref{extrapol-ulims} we compare photon fluxes, $N(>0.5\,\rm TeV)$, computed from the power-law fit (eq.~\ref{hawc-fit}) with the extrapolations of LAT spectra. HAWC photon fluxes relate to the normalizations $K$ through the integration of the differential spectra~(eqs.~\ref{nobs-ebl} and~\ref{hawc-fit}),
\begin{equation}
N_{obs}(>0.5~{\rm TeV}) = {4\sqrt{2}\over 3}\,K \cdot 1\, {\rm TeV}\cdot\,e^{-z/z_h}\,  . \label{phi1}
\end{equation} 
The (green, TeV=P) dots shown in Figure~\ref{extrapol-ulims} are for the objects with $TS>9$, while the rest are shown with the respective HAWC upper limit, using $K_{2\sigma}$ in expression~(\ref{phi1}). Thirteen objects have HAWC limits below, or close to, the LAT extrapolations. The VHE sources undetected by HAWC with limits below the LAT extrapolation are: IC~310, I~Zw~187 (1ES~1727+502), B3~2247+381, and 1ES~2344+514 (Figure~\ref{3fhl-xtrapol}). These are discussed in \S\ref{sec:undetected}. The HAWC limits for the TeV candidates PMN~J0816--1311, SBS~0812+578, and 1ES~2321+419 are also below the LAT extrapolations.

\subsubsection{Source classes}

\begin{deluxetable}{lcrrc}
\tablecaption{Average significances for different groups of sources. \label{stats-groups}}
\tablehead{ 
\colhead{Source} & \colhead{Number} & \multicolumn{2}{c}{Significances} & \colhead{p-value} \\
\colhead{group} & \colhead{of objects} &   \colhead{Mean} & \colhead{Std. dev}  & \colhead{\empty} \\ 
\colhead{\empty} & \colhead{$N$} & \colhead{$\mu(s)$} & \colhead{$\sigma_{s}$} & \colhead{$P\left[\mu(s)>x\right]$} }
\startdata
All            & 136 & +0.769 & 5.762 & $5.58\times 10^{-19}$ \\ 
All--Mkn & 134 & +0.175 & 1.209 & 0.022  \\ 
\hline
Starburst & 1 & +1.21 & \nodata & 0.113 \\
Radiogalaxies & 6 & +1.082 & 1.403 & $4.03\times 10^{-3}$ \\
BL Lacs--Mkn & 113 & +0.220 & 1.179 & $9.75\times 10^{-3}$ \\
FSRQ + fsrq & 6 & $-0.872$ & 0.609 & 0.984 \\
bcu & 8 & $-0.487$ & 1.028 & 0.916 \\
\hline
TeV = P (clean) & 30 & +0.768 & 1.228 & $1.29\times 10^{-5}$ \\
TeV = C (clean) & 32 & +0.124 & 1.152 & 0.241 \\
TeV = N (clean) & 72 & $-0.050$ & 1.142 & 0.665  \\
\enddata 
\end{deluxetable}

\begin{figure}
\includegraphics[width=\hsize]{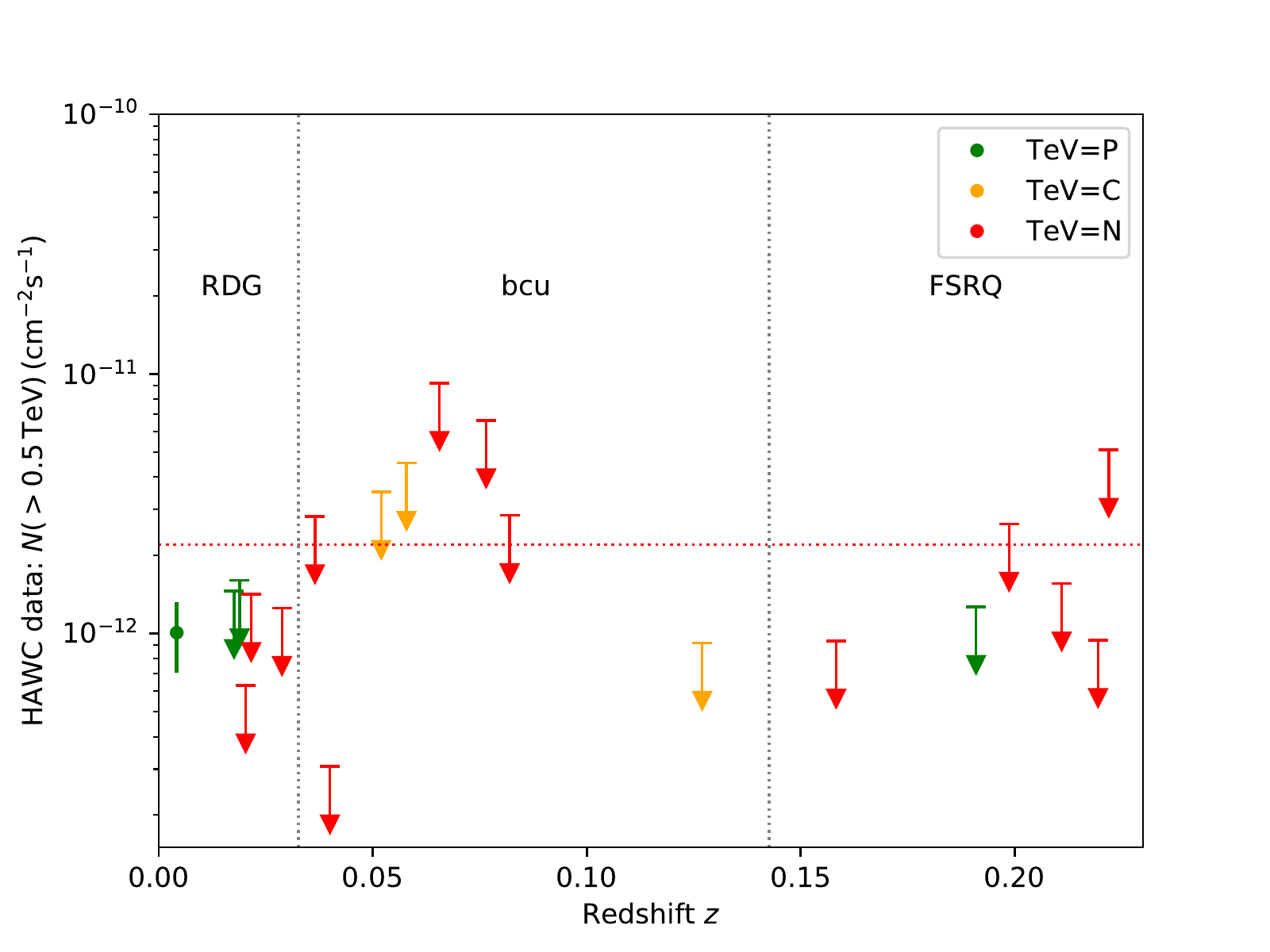}
\caption{HAWC photon flux upper limits for radiogalaxies (RDG), blazars candidates of uncertain type (BCU), and flat-spectrum radio quasars (FSRQ), naturally segregated in redshift integrals. The green point at the lowest redshift correspond to M87. Only three RDGs and one FSRQ have a TeV=P flag. 3C~264 was reported as a relatively faint VHE source after the release of the 3FHL catalog. The red dotted line is the 30~mCrab reference, while the vertical dotted lines separate the types of sources. \label{nobs-1}}
\end{figure}

\begin{figure}
\includegraphics[width=\hsize]{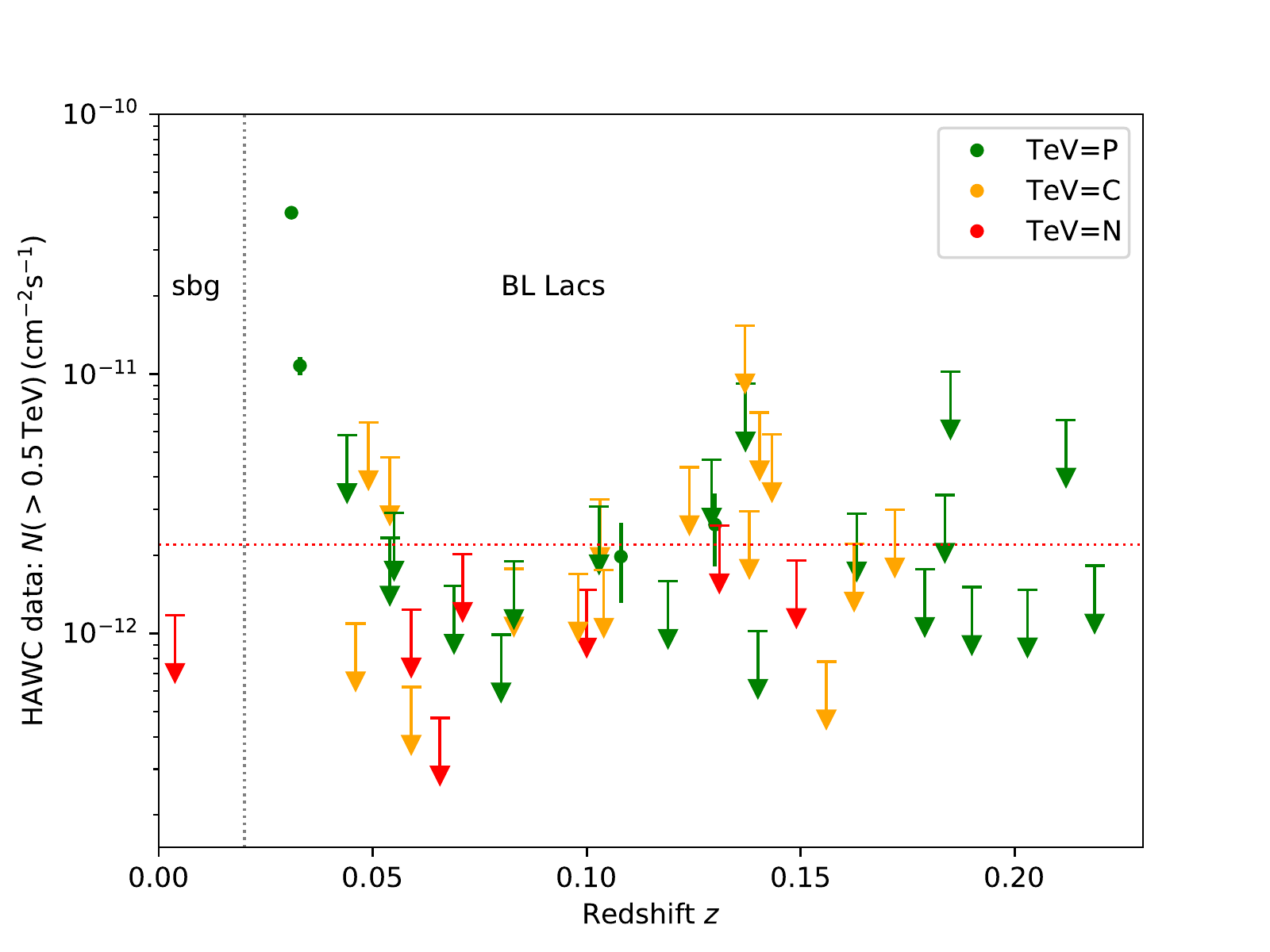}
\caption{HAWC photon flux upper limits for BL Lac objects and the starburst galaxy NGC~1068. Only BL Lac objects with LAT extrapolation above $2\times 10^{-13}\,\rm cm^{-2}s^{-1}$ are shown, as in Figure~\ref{3fhl-xtrapol}. The redshift interval is restricted to $z<0.23$, as in Figure~\ref{nobs-1}. No BL Lac is extrapolated above that flux beyond that distance. Four flux points are shown: Mkn~421 and Mkn~501 above $10^{-11}\,{\rm cm^{-2}s^{-1}}$; and VER~J0521+211 and 1ES~1215+303 slightly below and above the red line at redshifts 0.11 and 0.13, respectively. \label{nobs-2}}
\end{figure}

Table~\ref{stats-groups} presents average significances for distinct groups of sources. While GeV emission is known to occur in these sources, 80\% of the objects in our sample have not been detected in the VHE range. Furthermore, HAWC is testing long-term average emission, in contrast to the relatively brief IACTs observations. The group statistics, reinforced by the HAWC measurements as function of redshift for different source classes (Figures~\ref{nobs-1} and~\ref{nobs-2}), are briefly summarized as follows: \begin{itemize}
\item Except for the one starburst, radiogalaxies constitute the nearest class of sources in our sample (Figure~\ref{nobs-1}). All six are nearer than $z=0.03$, and happen to be at favorable declinations, from $\delta=+4^\circ$ (NGC 1278) to $\delta=+41^\circ$ (IC~310 and NGC~1275). Four of them are known VHE sources, including 3C~264, reported after the publication of the 3FHL, that shows a $+1.9\sigma$ excess in the HAWC data. As shown in Figure~\ref{nobs-1}, their bounds are at the $N(>0.5\,{\rm TeV})\sim 10^{-12}\,\rm cm^{-2}s^{-1}$ level. The mean significance for this group has a p-value of 0.4\%. 
\item The blazars candidates of uncertain type studied here are at redshifts intermediate between those of radiogalaxies and FSRQs. None of the bcu's studied here has been reported in the VHE regime. The HAWC upper limits shown in Figure~\ref{nobs-1} are not particularly constraining, owing to the unfavorable declinations of these sources, seven out of eight outside $+0^\circ \leq \delta\leq +50^\circ$.
\item Flat-spectrum radio quasars constitute the most distant class of source in our sample. Five out of six have not been detected as VHE sources, and do not comply with the LAT requirement for the TeV=C flag. They appear mostly as underfluctuations in the HAWC data. Figure~\ref{nobs-1} shows the upper limits on photon fluxes, most of them below the 30~mCrab level.
\item BL Lacertae objects constitute the clear majority of our sample, and of VHE sources. Still, 89 of the 117 BL Lac objects studied here have not been reported in the VHE regime. Figure~\ref{nobs-2} shows the HAWC fluxes for Mkn~421, Mkn~501, VER J0521+211 and 1ES~1215+303, at about 2 to $3\times 10^{-12}\,\rm cm^{-2}s^{-1}$ for the two weaker cases. 
\end{itemize}
The TeV=P flagged sources appear with a p-value at the $10^{-5}$ level, {excluding Mkn~421 and Mkn~501}, as expected for a sub-threshold persistent TeV emission in these known sub-TeV emitters. The TeV=C candidates and TeV=N groups do not provide any collective hint of emission. The p-values quoted do not account for the number of trials used in testing different groups; they are quoted as indicative of the potential presence of persistent TeV emission at levels $\lesssim 10^{-12}\,\rm cm^{-2}s^{-1}$. 

\subsection{Spectral fits for the most significant sources\label{subsec:optimized}}
\begin{figure}
\includegraphics[width=\hsize]{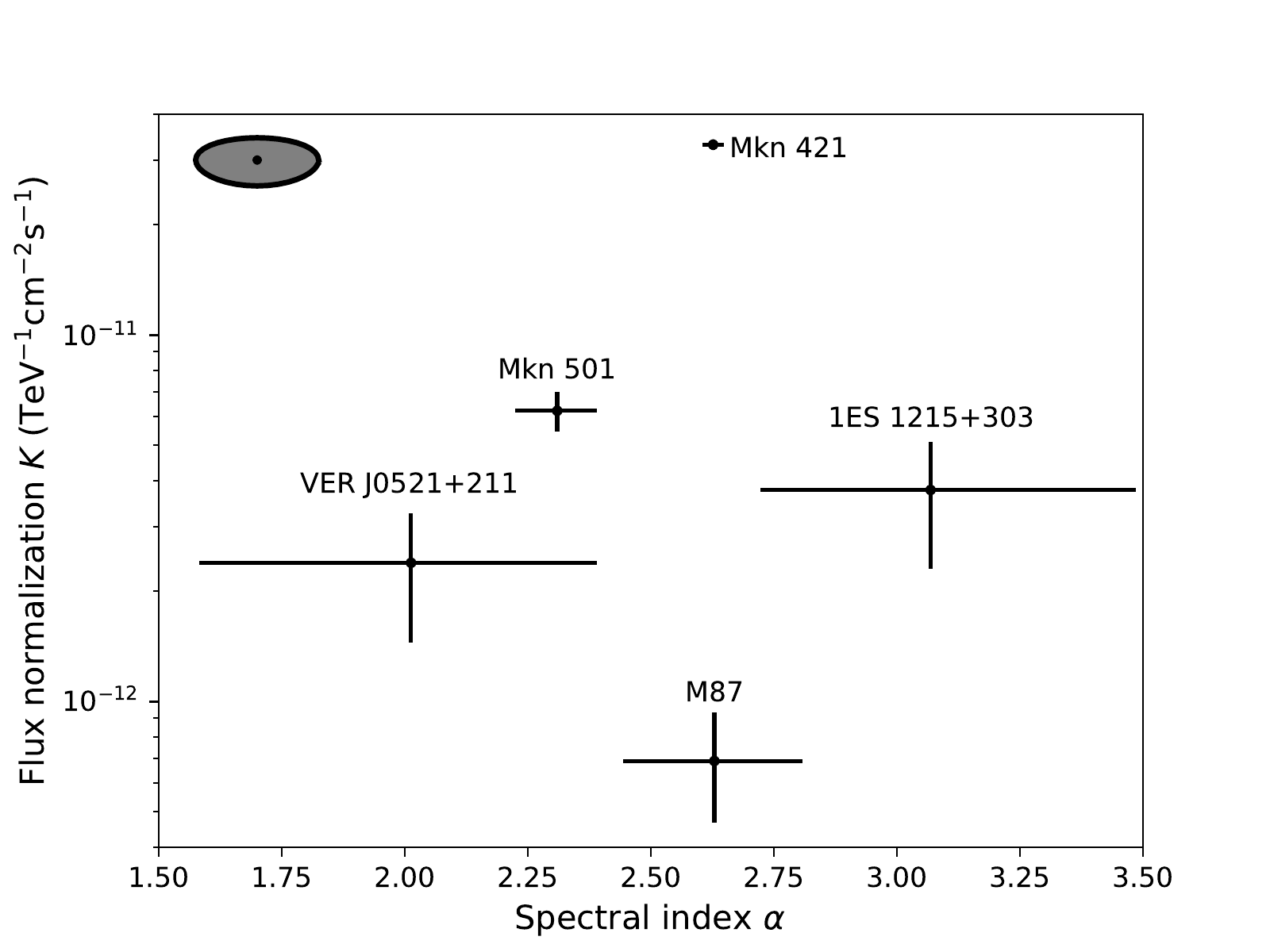}
\caption{Optimized spectral fits for five selected sources. Mkn~421 and Mkn~501, the two highest points, present the best statistics. VER~J5021+211 and 1ES~1215+303 have the extreme spectral indices, although with a larger uncertainty. M87 appears with the smallest normalization and a spectral index consistent with $\alpha=2.5$. The black and grey oval on the upper left represents systematic uncertainties of 15\% in the 1~TeV normalization ($K$) and 5\% on spectral index ($\alpha$).
\label{spectral-fits}}
\end{figure}

We computed optimized spectra for the $TS>9$  sources, fitting together normalizations and spectral indices. The fits are summarized in Figure~\ref{spectral-fits}, together with the systematic uncertainties. These have been quantified as 15\% in $K$, the 1~TeV normalization, and 5\% in $\alpha$, the spectral index. A short discussion on each of these five sources follows.

\subsubsection{Markarian 421}

\begin{figure}
\includegraphics[width=\hsize]{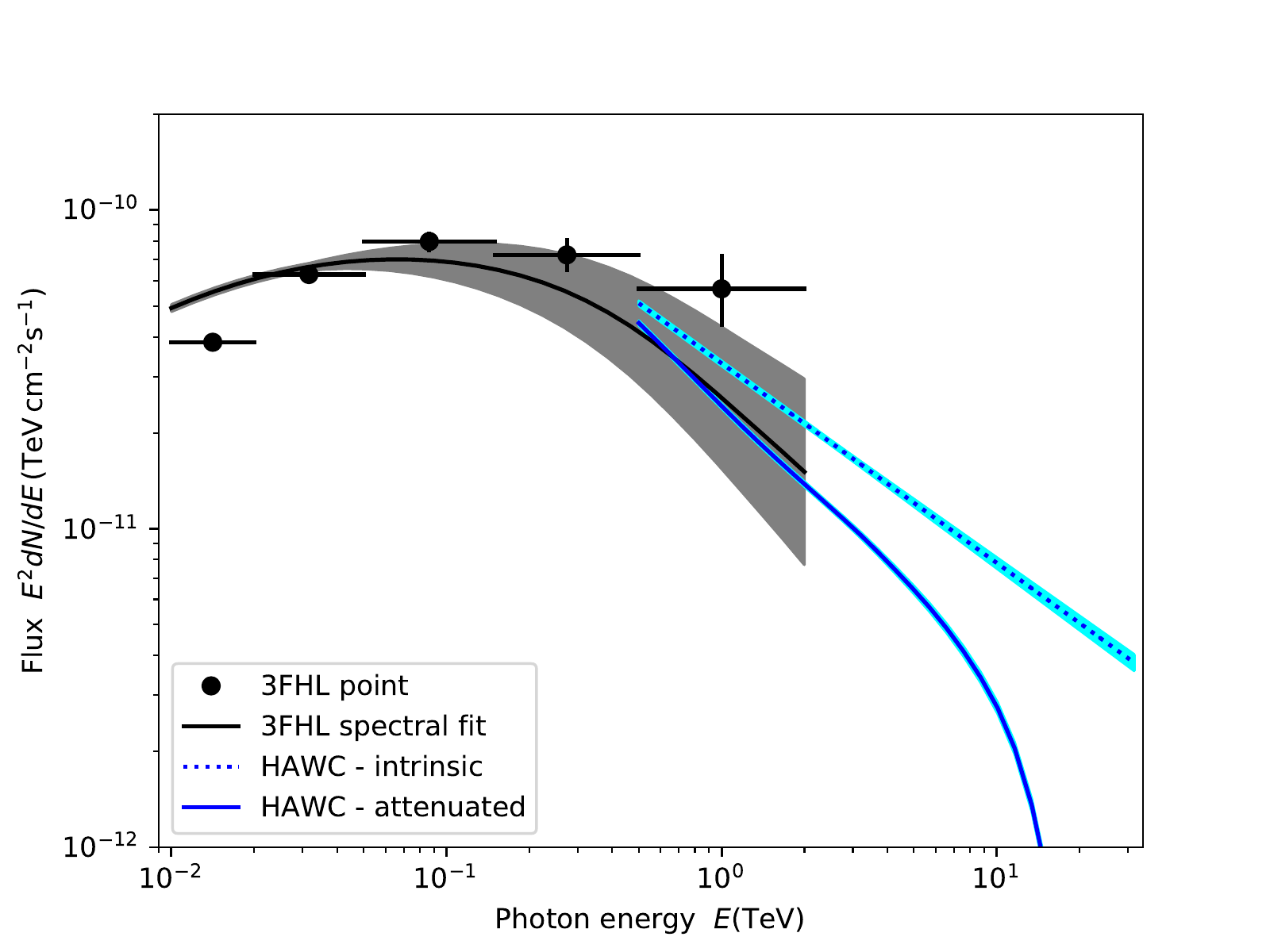}
\caption{{\bf Spectra} of Mkn~421 measured by LAT, including  EBL attenuation (black and grey, on the left) and by HAWC (blue, on the right). The full-dotted line represents the intrinsic spectrum, with statistical errors, and the full line the fit with EBL attenuation. The two attenuated fits match well at their (0.5-2.0)~TeV intersection. 
\label{mkn421-spectrum}}
\end{figure}

Markarian~421 is the brightest persistent extragalactic object in the TeV sky. For the default $\alpha=2.5$ search we computed ${TS}=4167$ for $K=(29.5\pm 0.5_{\rm stat} \pm 4.4_{\rm syst})\times 10^{-12}\,\rm TeV^{-1}cm^{-2}s^{-1}$. The optimized power-law fit for the intrinsic spectrum resulted in,
\begin{equation}
{dN\over dE} = (33.0\pm 0.6_{\rm stat} \pm 4.9_{\rm syst})\times 10^{-12}\, \left(E \over 1\,{\rm TeV}\right)^{-2.63\pm 0.02_{\rm stat} \pm 0.13_{\rm syst}} {\rm TeV^{-1}cm^{-2}s^{-1}} \, ,
\end{equation}
with $TS=4193$, i.e. an increase in the test-statistic $\Delta TS=26$ relative to $\alpha$ fixed at 2.5. The integration to the observed photon flux, using $z_h=0.116$ for $\alpha=2.63$ in eq.~\ref{nobs-ebl}, gives $N_{\rm obs}=(48\pm 8)\times 10^{-12}\,\rm cm^{-2}s^{-1}$, somewhat larger than the $\alpha=2.5$ estimate shown in Figure~\ref{extrapol-ulims}. The {\em intrinsic} energy flux, $f_E=(132\pm 22)\times 10^{-12}\,\rm erg\, cm^{-2}s^{-1}$, translates into a luminosity per solid angle of $f_E d_L(z)^{2}= L(>0.5\,{\rm TeV})/\Delta\Omega = (2.1\pm 0.4)\times 10^{43} \,\rm erg~s^{-1} sr^{-1}$, about one third (32\%) of the (10~GeV-1~TeV) luminosity per steradian inferred from the {\em Fermi}-LAT measurements (Figure~\ref{muestra-z-lumi}).The HAWC spectrum is consistent with IACT observations made by MAGIC between 2007 and 2009 \citep{2016A&A...593A..91A}. Using an energy threshold of 400~GeV, these authors found photon fluxes varying from $N_{\rm min}(>0.5~{\rm TeV}) = 9.33\times 10^{-12}\,\rm cm^{-2}s^{-1}$, to $N_{\rm max}(>0.5~{\rm TeV}) = 2.22\times 10^{-10}\,\rm cm^{-2}s^{-1}$, weakly dependent on their assumed differential index of 2.5.

{The 3FHL catalog and HAWC spectra for Mkn~421 are shown together in Figure}~\ref{mkn421-spectrum}. There is a fairly good match between the two fits, with the curves intersecting at about 0.66~TeV. The local LAT spectral index at 1~TeV, $2.5\pm 0.2$, is consistent within uncertainties with the HAWC spectral index. IACT observations between 100~GeV and 5~TeV, performed around 2005 with MAGIC, resulted in a spectral index of $(2.20\pm 0.08)$, with indications of a cut-off in the intrinsic spectrum~\citep{2007ApJ...663..125A}. A detailed spectral analysis of Mkn~421, and Mkn~501, with HAWC data will be presented in a separate publication. Preliminary results can be found in~\citet{2019ICRC...36..654C}.

\subsubsection{Markarian 501}

\begin{figure}
\includegraphics[width=\hsize]{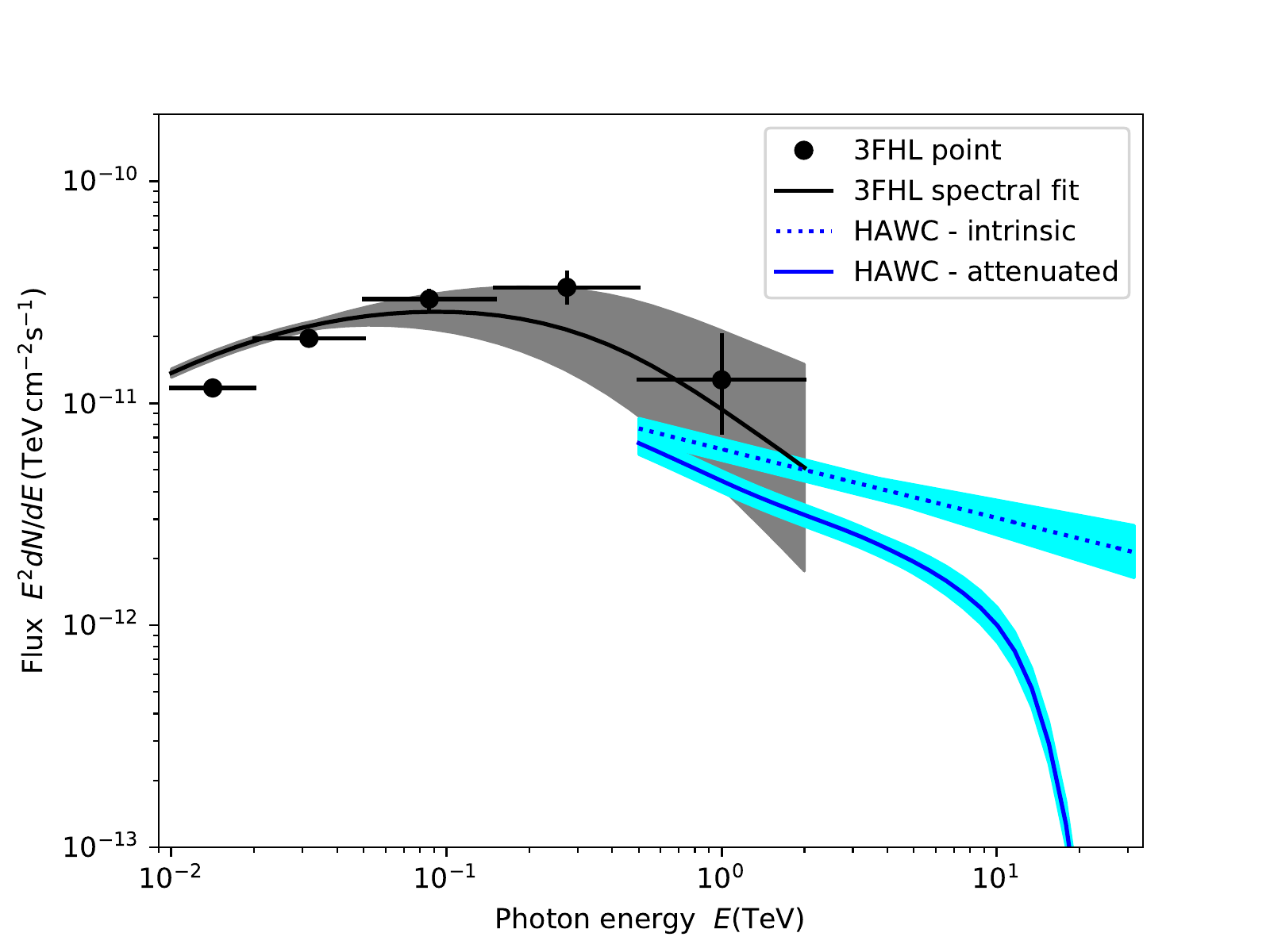}
\caption{Spectrum of Mkn~501 measured by LAT and EBL attenuated  (black and grey) and by HAWC (blue). The full-dotted line represents the HAWC intrinsic spectrum, with statistical errors, and the full line the fit with EBL attenuation. 
\label{mkn501-spectrum}}
\end{figure}

The TeV emission from Markarian~501 observed by HAWC is not as steady as that of Mkn~421~\citep{2017ApJ...841..100A}. In fact, the statistical significance of the time-averaged TeV emission of Mkn~501 has {\em decreased} with increased HAWC exposure. The $\alpha=2.5$ search resulted in a test-statistic ${TS}=276.97$ for $K=(7.74\pm 0.49_{\rm stat}\pm 1.16_{\rm syst})\times 10^{-12}\,\rm TeV^{-1}cm^{-2}s^{-1}$. The optimized power-law fit is,
\begin{equation}
{dN\over dE} = (6.21\pm 0.69_{\rm stat} \pm 0.93_{\rm syst})\times 10^{-12}\, \left(E \over 1\,{\rm TeV}\right)^{-2.31\pm 0.08_{\rm stat} \pm 0.12_{\rm syst}} {\rm TeV^{-1}cm^{-2}s^{-1}} \, , \label{mkn501-fit}
\end{equation}
with $TS=280.28$, representing a moderate increase in the test-statistic $\Delta TS=3.33$. The integrated observed photon flux is $N_{\rm obs}=(8.5\pm 2.6)\times 10^{-12}\,\rm cm^{-2}s^{-1}$, accounting for EBL attenuation. The {\em intrinsic} energy flux, $f_E=(40\pm 16)\times 10^{-12}\,\rm erg\, cm^{-2}s^{-1}$, translates into a luminosity per solid angle of $d_L(z)^{2}f_E = L(>0.5\,{\rm TeV})/\Delta\Omega = (7.3\pm 2.8)\times 10^{42} \,\rm erg~s^{-1} sr^{-1}$, which is about 25\% of the (10~GeV-1~TeV) luminosity per steradian measured by {\em Fermi}-LAT (Figure~\ref{muestra-z-lumi}). This fraction is similar to that observed for Mkn~421.

The 3FHL catalog and HAWC spectra for Mkn~501 are shown together in Figure~\ref{mkn501-spectrum}. The agreement is not as good as for Mkn~421: the Mkn~501 spectrum is harder and lies below the measurement by the LAT. The local LAT spectral index at 1~TeV is $2.58\pm 0.35$, just consistent with HAWC when accounting for the propagation of the uncertainty in the curvature parameter $\beta$ of the 3FHL fit. We note that Mkn~501 has a variability index $V_{\rm bayes}=4$ in the 3FHL catalog.

The literature on the VHE characteristics of Mkn~501 is extensive. Its high TeV variability was noted shortly after its 1996 discovery, when the HEGRA group reported flux variations of an order of magnitude observed in mid-1997~\citep{1996ApJ...456L..83Q,1997A&A...327L...5A}. Contemporaneous 10~m Whipple data confirmed the high state, adding indications of a curved spectrum favored over a simple power-law~\citep{1998ApJ...501L..17S}.  Further monitoring showed strong variations in flux, although with stable spectra best-fitted by a power-law plus an intrinsic cut-off at around 6~TeV~\citep{1999A&A...349...11A}. Observations in 1998-1999 showed lower activity, with evidence for spectral curvature and steepening~\citep{2001ApJ...546..898A}. Mkn~501 observations renewed in 2005, when the MAGIC telescope measured strong and very fast variability, with spectral indices ranging from $\sim 2.0$ to $2.7$~\citep{2007ApJ...669..862A}. Data taken the following year (2006) by MAGIC permitted to characterize a low activity state, with fluxes similar to those reported by VERITAS and MAGIC from data taken in the 2009 joint observations with the {\em Fermi}-LAT~\citep{2009ApJ...705.1624A,2011ApJ...729....2A}. The contemporaneous LAT data showed spectral variability also present in the GeV range, with index variations $\Delta\alpha\sim 1$ during the first 480~days of {\em Fermi} observations~\citep{2011ApJ...727..129A}. 

In the last decade, EAS arrays have been able to reach the sensitivities needed to perform long term monitoring of Mkn~501: the ARGO collaboration reported variations of a factor of six in flux observed between October 2011 and April 2012~\citep{2012ApJ...758....2B}. The HAWC collaboration presented light curves for the Crab, Mkn~421 and Mkn~501 from its first 17~months of observations~\citep{2017ApJ...841..100A}. The HAWC light curve of Mkn~501 showed a low flux baseline with a handful of very short strong flares. The simple power-law fit (of index $2.84$) was disfavored against a power-law with exponential cut-off. 

The variability of Mkn~501 has prevented a baseline characterization of this object. In \citet{2019ICRC...36..654C} we presented the HAWC spectrum of Mkn~501 using data acquired between June 2015 and December 2017. Even though spectra with exponential cuts were tested, the Mkn~501 data proved consistent with a pure power-law of index~$2.40\pm 0.06$. While there might be a slight decrease in the spectral index ($2.31\pm 0.08 \pm 0.20$), the flux reported here is about half that reported in~\citet{2019ICRC...36..654C}. The four year average TeV flux observed by HAWC is a factor of two above the lowest activity observed so far~\citep{2009ApJ...705.1624A,2011ApJ...729....2A}.

\subsubsection{M87}

\begin{figure}
\includegraphics[width=\hsize]{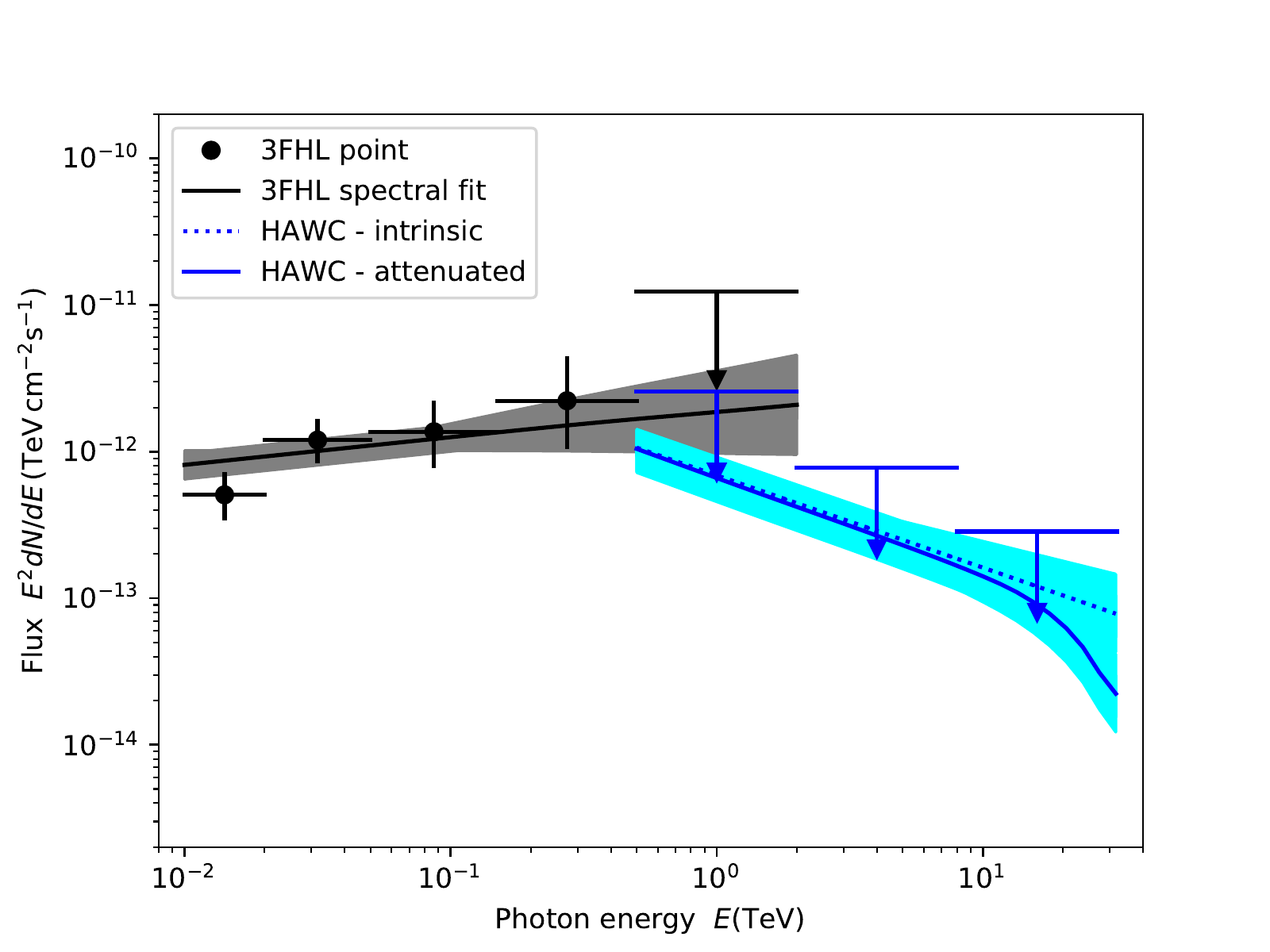}
\caption{High energy spectra of M87 from the LAT Catalog and HAWC observations. The data suggest a steepening of the spectrum at energies $\sim 1~\rm TeV$. The HAWC optimized power-law fit has $TS=13.2$. The quasi-differential computation for the (2.0-8.0)~TeV interval is at $TS=13.1$. 
\label{m87-espectro}}
\end{figure}

M87 is the central galaxy of the Virgo cluster, a giant elliptical at a distance of just $(16.4\pm 0.5)~\rm Mpc$, as measured independently of its redshift, $z=0.0042$~\citep{2010A&A...524A..71B}. With an optical magnitude $V=8.6$ and an angular diameter of about 8~arcmin, M87 has been imaged in detail for over a century, showing a distinct bright active nucleus and a single optical jet~\citep{1918PLicO..13....9C,1998ApJ...493L..83T}. The mass of the central black hole has been measured to be $(6.5\pm 0.7)\times 10^{9}\,M_\odot$, through its imaging with the Event Horizon Telescope~\citep{2019ApJ...875L...1E}. 

Also known as Virgo~A, it is a bright object all throughout the electromagnetic spectrum. As 3FHL~J1230.8+1223, it is a $12.1\sigma$ detection above 10~GeV as reported in the 3FHL catalog, with $4.8\sigma$ in 150-500~GeV, and a $2\sigma$ upper limit in the top (0.5-2.0)~TeV band. M87 has been observed frequently in the VHE regime since its 2003 discovery by the HEGRA collaboration~\citep{2003A&A...403L...1A}. The temporal behavior of M87 in the high-energy and VHE bands has been reviewed by~\citet{2019A&A...623A...2A}. Fluxes can vary by a factor of ten between low, mid and high states, with indications of the spectral index varying from $2.6$ in low states, to $2.2$ in high states. In addition to the low and high states, time variability on single-day timescales during high states has been reported by the H.E.S.S. and MAGIC collaborations~\citep{2007AIPC..921..147B,2008ApJ...685L..23A}. 

The default HAWC search at the M87 location gave a test-statistic of $TS=12.93$ for a normalization $K=(0.56\pm 0.16_{\rm stat}\pm 0.08_{\rm syst})\times 10^{-12}\,\rm TeV^{-1}cm^{-2}s^{-1}$.  The optimized power-law fit for the intrinsic spectrum is,
\begin{equation}
{dN\over dE} = (0.69\pm 0.22_{\rm stat} \pm 0.10_{\rm syst})\times 10^{-12}\, \left(E \over 1\,{\rm TeV}\right)^{-2.63\pm 0.18_{\rm stat} \pm 0.13_{\rm syst}} {\rm TeV^{-1}cm^{-2}s^{-1}} \, ,
\end{equation}
with $TS=13.19$, i.e. a non-significant increase in the test-statistic, $\Delta TS=0.26$, showing $\alpha=2.5$ to be an acceptable solution within the statistics of the optimized fit. 
The integrated photon flux is $N_{\rm obs}(>0.5~{\rm TeV})=(1.3\pm 0.9)\times 10^{-12}\,\rm cm^{-2}s^{-1}$, with less than 4\% of attenuation by the EBL. The energy flux, $f_E=(2.7\pm 2.4)\times 10^{-12}\,\rm erg\,cm^{-2}s^{-1}$, translates into a luminosity per solid angle of $L(>0.5~{\rm TeV})/\Delta\Omega = (6.9\pm 6.3)\times 10^{39}\,\rm erg\,s^{-1}sr^{-1}$, which is about 25\% of that in the 10~GeV-1~TeV range, from the respective 3FHL parameters. The lower apparent power with respect to the two nearest BL Lacs is attributed to the off-axis viewing of the jet. Still, the relative power when compared to the LAT regime appears to be similar.

From \citet{2019A&A...623A...2A}, we computed photon fluxes from M87 observations by IACTs, in different activity states:
\begin{itemize}
\item High state: $N(>0.5\,{\rm TeV}) = (5.74^{+1.14}_{-1.47})\times 10^{-12}\,\rm cm^{-2}s^{-1}$,
\item Mid state: $N(>0.5\,{\rm TeV}) = (1.39^{+0.47}_{-0.43})\times 10^{-12}\,\rm cm^{-2}s^{-1}$,
\item Low state: $N(>0.5\,{\rm TeV}) = (2.85^{+2.07}_{-1.53})\times 10^{-13}\,\rm cm^{-2}s^{-1}$.
\end{itemize}
The 4.5~year averaged emission, as indicated by the HAWC data, matches the mid state, with a relatively steep spectral index. The comparison with the 3FHL data, shown in Figure~\ref{m87-espectro}, points to a steepening in the spectrum. While the LAT and HAWC data are not contemporaneous, with a variability index $V_{\rm bayes}=1$ M87 does not stand as a variable in the 3FHL catalog. Still, an analysis of joint contemporaneous LAT-HAWC data is desirable.

\subsubsection{VER J0521+211 \label{ver_j0521+211}}

\begin{figure}
\includegraphics[width=\hsize]{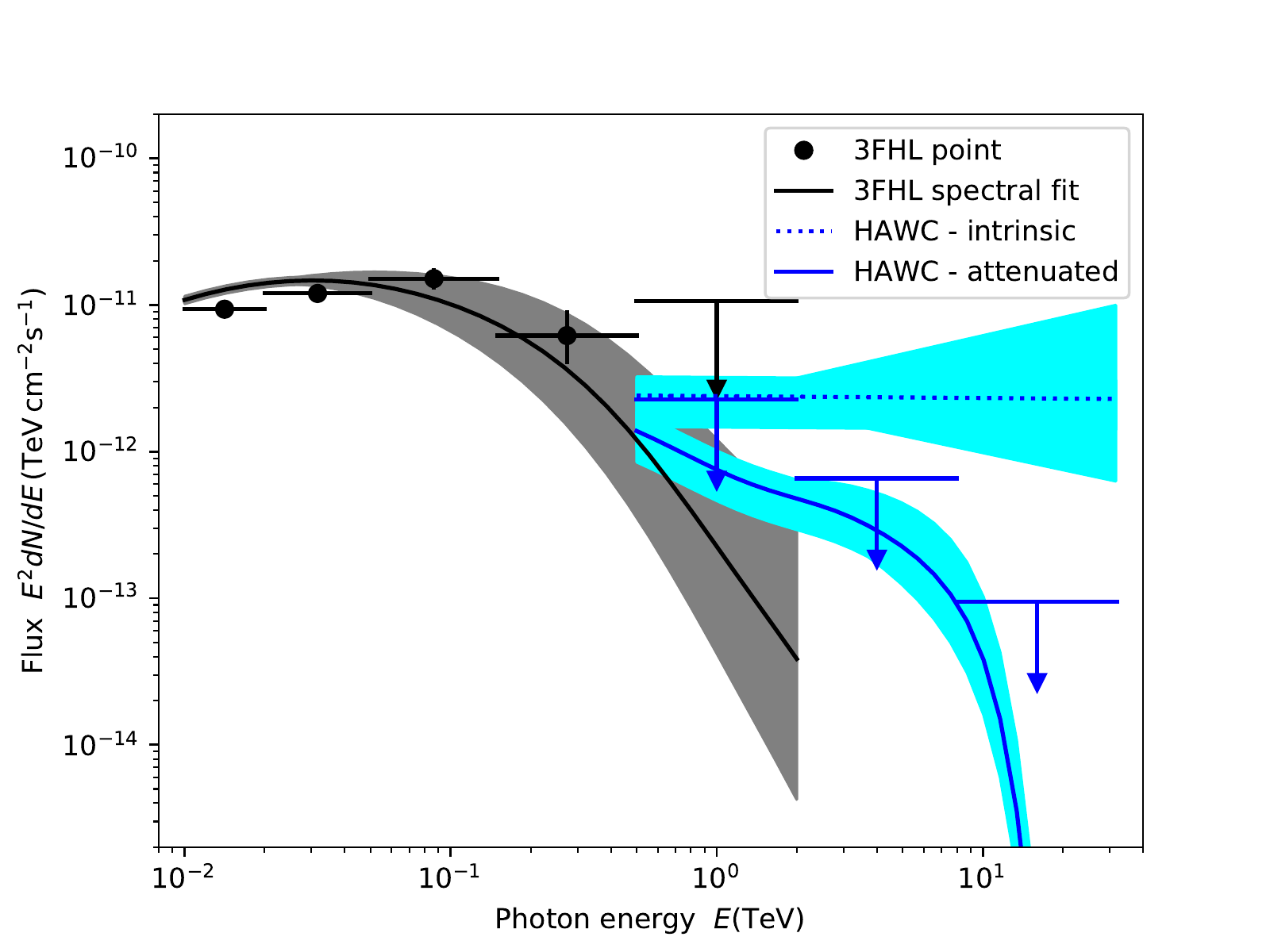}
\caption{Spectral fits of VER~J0521.7+2112 from {\em Fermi}-LAT (grey) and HAWC (blue) data. The best fit of the HAWC data is for an intrinsic power-law of index 2.0, which remains harder than the 3FHL log-parabola fit after considering EBL attenuation at the assumed redshift. The HAWC quasi-differential limit at (2.0-8.0)~TeV is for a $TS=9.1$.\label{verj0521-espectro}}
\end{figure} 

VER~J0521+211 was discovered as a TeV $\gamma$-ray source by the VERITAS collaboration, during observations following-up its detection above 30~GeV by LAT~\citep{2009ATel.2260....1O}. 3FHL~J0521.7+2112 itself is associated to the radio-loud BL Lac object TXS~0518+211, and the X-ray {\it ROSAT} source RX~J0521.7+2112. The redshift survey of \citet{2013ApJ...764..135S} assigned $z=0.108$ to the optical counterpart. This is the value listed in the 3FHL catalog and assumed for this analysis. \citet{2013ApJ...776...69A} did not confirm this redshift, and neither did \citet{2017ApJ...837..144P}, who only set a lower limit $z>0.18$ using deep spectroscopy with the GTC 10.4~m. The VHE emission from this object has been measured at least up to 1~TeV, with no apparent decline in the spectrum~\citep{2015ICRC...34..864P}. This makes the high-energy characterization of VER~J0521+211 of particular interest.

The $\alpha=2.5$ search resulted in $TS=9.49$ for $K=(2.85\pm 0.93_{\rm stat} \pm 0.43_{\rm syst})\times 10^{-12}\,\rm TeV^{-1}cm^{-2}s^{-1}$. The best power-law fit for the intrinsic spectrum is,
\begin{equation}
{dN\over dE} = (2.39\pm 0.89_{\rm stat} \pm 0.36_{\rm syst})\times 10^{-12}\, \left(E \over 1\,{\rm TeV}\right)^{-2.01\pm 0.38_{\rm stat} \pm 0.10_{\rm syst} } {\rm TeV^{-1}cm^{-2}s^{-1}} \, ,
\end{equation}
with $TS=10.34$, representing a modest increase in the test-statistic $\Delta TS=0.85$ with respect to the default search. This is the hardest AGN spectrum for the $TS>9$ HAWC sample. The integrated observed photon flux is $N_{obs}(>0.5~{\rm TeV})=(1.5\pm 1.1)\times 10^{-12}\,\rm cm^{-2}s^{-1}$, attenuated by a factor of 2/3 due to the EBL. The energy flux and luminosity per solid angle cannot be accurately determined with the spectral index so close to $2.0$. If we take $E^{2}dN/dE$ at 1~TeV as indicative, we get an estimated $f_E\sim (3.8\pm 1.4)\times 10^{-12}\,\rm erg\,cm^{-2}s^{-1}$ and $d_L^{2}f_E\sim (8.4\pm 3.1)\times 10^{42}\,\rm erg\,s^{-1}sr^{-1}$. Notwithstanding the uncertain distance, this source appears particularly luminous in the GeV regime, with $L(10~{\rm GeV}-1~{\rm TeV})/\Delta\Omega \sim 1.4\times 10^{44}\,\rm erg\,s^{-1}sr^{-1}$, standing above the two Markarians in Figure~\ref{muestra-z-lumi}.

The 3FHL data, shown together with the HAWC spectral fit in Figure~\ref{verj0521-espectro}, has strong detections up to the 150-500~GeV band, with a highest energy photon of 370~GeV. The 3FHL fit is a log-parabola, transiting from a hard spectrum at about 30~GeV to a very steep local spectral index of $3.7\pm 0.7$ at 1~TeV (4.3 when attenuating the 3FHL spectrum). The attenuated HAWC spectrum corresponds to an observed spectral index of $\sim 2.7$, mostly inconsistent with the LAT fit. The data may be reconciled through an intrinsic hardening at about 200 or 300~GeV. The HAWC quasi-differential bound in the 0.5-2.0~TeV band is a factor of 4.6 lower than the 3FHL limit in the same band. We note that the quasi-differential analysis gave $TS=9.1$ for the 2.0-8.0~TeV band, optimal in terms of the HAWC instrumental response. With the assumed redshift, $\tau$ is due to range from 1.6 to 3.2 in that energy interval, pointing to an intrinsically hard spectrum. The analysis gave $TS=0$ for 8.0-32.0~TeV, expected to be heavily attenuated for the assumed redshift. Additional HAWC data analysis should allow to further constrain the shape of the TeV spectrum VER~J0521+211, testing the redshift assumption. {We also note that VER~J0521+211 has a variability index $V_{\rm bayes}=4$ in the 3FHL catalog, indicating that a joint analysis of contemporaneous LAT and HAWC data would be relevant.}

{As mentioned in \S\ref{subsec:sample}, VER~J0521+211 is located $3.07^\circ$ from the Crab Nebula, the brightest source in the 2HWC and 3HWC catalogs. This angular distance corresponds to three times the 68\% containment angle $\psi_{68}$ for ${\cal B}=1$, and $> 6\psi_{68}$ for ${\cal B}>2$. We tested for potential contamination repeating the Maximum-Likelihood test with $\alpha=2.5$ at five locations equidistant from the Crab Nebula, forming together with  VER~J0521+211 an hexagon around the Crab. These provided test-statistics $TS$ between $-4.46$ and $+1.54$, in contrast with $TS=+9.49$ at the location of VER~J0521+211.}

\subsubsection{1ES~1215+303}

\begin{figure}
\includegraphics[width=\hsize]{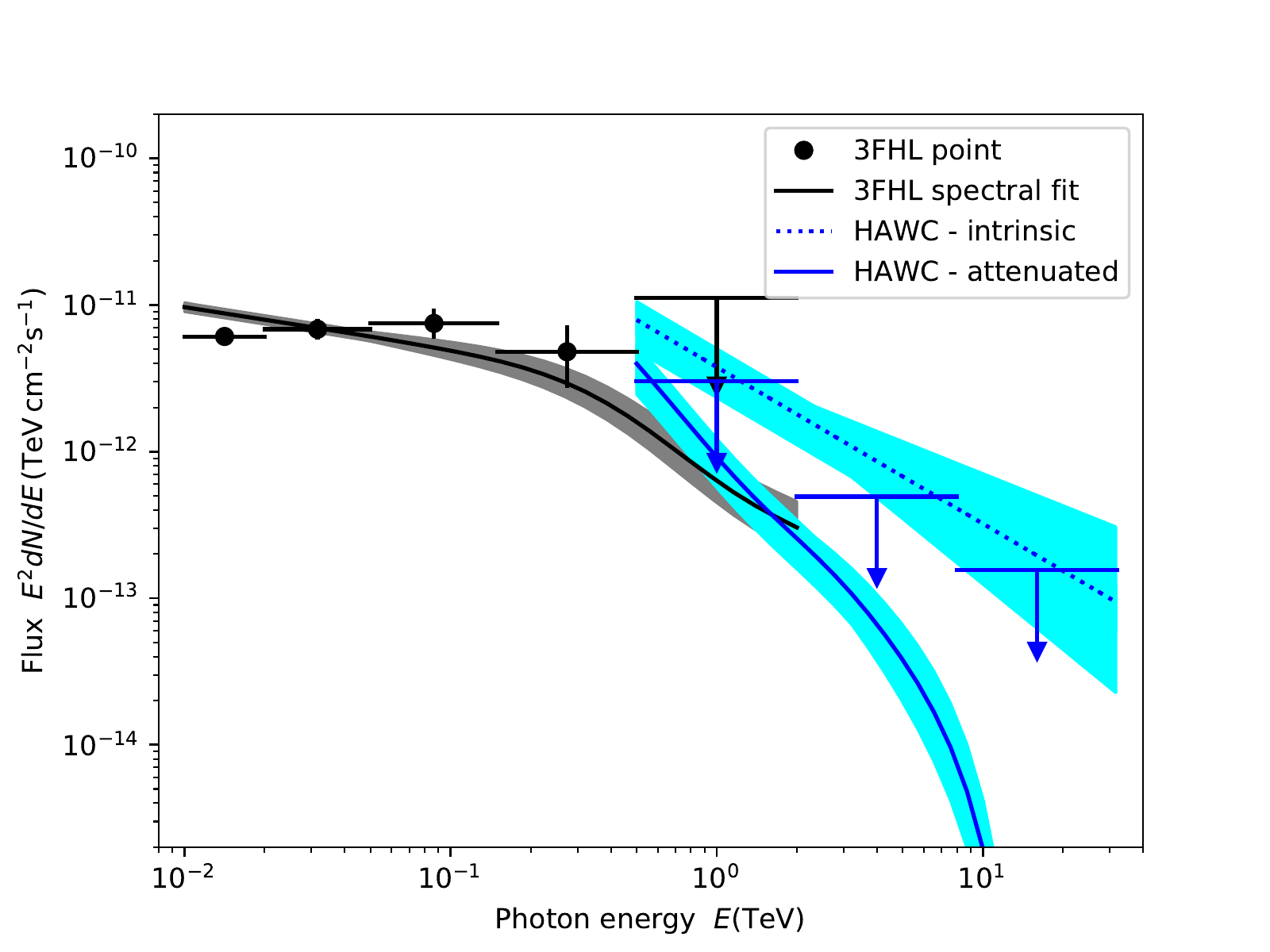}
\caption{High energy spectra of 1ES~1215+303 from the 3FHL catalog and HAWC observations. We note the intersection of both spectra at $\gtrsim 1\,\rm TeV$ and its decay, as it suffers EBL attenuation. This is the most distant source detected by HAWC so far.\label{1es1215-espectro}}
\end{figure}

1ES~1215+303 is one of six $\gamma$-ray emitting BL Lac objects located in the Northern part of the Coma Berenices constellation, five of them known to be VHE sources. The redshift of this HBL is now confirmed to be $z=0.130$, discarding the early measurement $z=0.237$ by~\citet{1993ApJS...84..109L}. The lower value was confirmed through optical spectroscopy at the GTC and, more recently, through the direct identification of a Ly$\,\alpha$ emission line~\citep{2017ApJ...837..144P,2019AJ....157...41F}. Catalogued as 3FHL~J1217.9+3006,  this object is well detected up to 500~GeV, and modeled with a power-law of index $\alpha=2.3\pm 0.1$. It has a variability index $V_{\rm bayes}=2$.

The default HAWC search gave $TS=11.36$ for $K=(4.64\pm 1.38_{\rm stat} \pm 0.70_{\rm syst} )\times 10^{-12}\,\rm TeV^{-1}cm^{-2}s^{-1}$, as the intrinsic normalization. The optimized power-law fit for the intrinsic spectrum is rather softer,
\begin{equation}
{dN\over dE} = (3.78\pm 1.36_{\rm stat} \pm 0.57_{\rm syst} )\times 10^{-12}\, \left(E \over 1\,{\rm TeV}\right)^{-3.07\pm 0.37_{\rm stat} \pm 0.15_{\rm syst} } {\rm TeV^{-1}cm^{-2}s^{-1}} \, ,
\end{equation}
with $TS=12.80$, representing a test statistic increase $\Delta TS=1.44$. The attenuated photon flux is $N_{\rm obs}(>0.5\,{\rm TeV})=(2.53\pm 1.35)\times 10^{-12}\,\rm cm^{-2}s^{-1}$. The integrated energy flux results in a luminosity per solid angle of $L(>0.5~{\rm TeV})/\Delta\Omega = (3.9\pm 2.7)\times 10^{43}\,\rm erg\,s^{-1}sr^{-1}$, given the luminosity distance of 585~Mpc. This is about 30\% of the corresponding value between 10~GeV and 1~TeV. The 3FHL and HAWC spectra, compatible within the uncertainties, are shown in Figure~\ref{1es1215-espectro}.

1ES~1215+303 was first detected as a VHE source by MAGIC in 2011~\citep{2011arXiv1110.6786L,2012A&A...544A.142A}. {It is usually observed together with PG~1218+304, located just 0.88 degrees away in the sky, for which we obtained a test statistic $TS=+2.24$. Given that both sources are known VHE emitters, the distance is not large enough to be certain that there is no overlap between both sources. With a redshift  $z=0.184$, PG~1218+304 is more distant and prone to be heavily attenuated above 1~TeV. A dedicated study with improved HAWC analysis tools is now pending.}

Long-term monitoring of this source by VERITAS was presented by~\citet{2013ApJ...779...92A}, prior to the report of a single short and intense flare seen in 2014 with VERITAS and {\em Fermi}-LAT~\citep{2017ApJ...836..205A}. During this episode the VHE flux of this HBL reached 2.4 times that of the Crab Nebula, with a variability timescale $\lesssim 3.6\,{\rm hours}$. The MAGIC spectrum of~\citet{2012A&A...544A.142A}, ranging from 70~GeV to 1.8~TeV, had an intrinsic spectral index of $2.96$, de-attenuated with the model of~\citet{2011MNRAS.410.2556D},  used in here. The MAGIC and HAWC spectral indices are in good agreement. The long-term joint monitoring of this source by LAT and VERITAS, spanning more than ten years, confirms the spectral index and points to four strong and short flares that occurred during full HAWC operations by~\citet{2020ApJ...891..170V}. We checked the dates of these four episodes and did not find evidence for them in the HAWC data. An optimized time-dependent analysis of this source with HAWC, beyond the scope of this paper, will be subject of future work.

\subsection{Limits on sources of interest \label{sec:undetected}}

\begin{deluxetable*}{llcccccc}
\tablecaption{Upper limits on 3FHL AGN candidate targets. \label{undetected}}
\tablehead{ \colhead{3FHL entry} & \colhead{Source} & \colhead{z} & \colhead{$\pm\sqrt{TS}$}  & \colhead{$K_{2\sigma}$} & \colhead{$N_{0.5}^{x}$} & \colhead{$N_{0.5}^{UL}$} } 
\startdata
\hline
3FHL J0214.5+5145  & TXS~0210+515 & 0.049 & $+1.62$ &  5.43 & 2.40 & 6.50 \\ 
3FHL J0316.6+4120   & IC~310  & 0.019 & $+0.86$  &  1.01 & 4.37 & 1.60 \\ 
3FHL J1428.5+4240 &  H~1426+428  & 0.129  & $+1.26$   &  8.18 & 1.21 & 4.67 \\
3FHL J1728.3+5013 &  I~Zw~187   & 0.055  & $-0.22$     &  2.57 & 3.25 & 2.91 \\
3FHL J2250.0+3825 & B3~2247+381  & 0.119 & $-0.27$ &  2.54 & 2.37 & 1.59 \\
3FHL J2347.0+5142 & 1ES~2344+514 & 0.044 & $+1.45$ &  4.62  & 7.12 & 5.80 \\\hline
\enddata
\tablecomments{$K_{2\sigma}$, the spectrum normalization at 1~TeV, is in units of $10^{-12}\,\rm TeV^{-1}cm^{-2}s^{-1}$. $N_{0.5}$ are extrapolated photon fluxes above 0.5~TeV and the corresponding upper limit, in units of $10^{-12}\,\rm cm^{-2}s^{-1}$.}
\end{deluxetable*}

The extrapolation of 3FHL catalog spectra (Figure~\ref{3fhl-xtrapol}) and of IACT spectral fits allowed us to identify nine AGN targets for potential HAWC detection (\S\ref{extrapol}). We presented evidence of persistent TeV emission for three of those targets, Mkn~421, Mkn~501, and M87, and for two sources below the 30~mCrab reference limit (marked as a red line in Figure~\ref{3fhl-xtrapol}), VER J0521+211 and 1ES~1215+303. On the other hand, six of our target sources (IC~310, 1ES~2344+514, TXS~0210+515, 1ES~1727+502, B3~2247+381, H~1426+428) were not detected. As shown in Figure~\ref{ulims-fit}, the HAWC sensitivity is dependent on source declination. We note that while the sources with $\sqrt{TS}>9$ are all in the range $+12^\circ<\delta<+40^\circ$, five of the undetected candidates are either North of $\delta=+50^\circ$, or farther than $z=0.1$. Declination is of particular relevance in here, as the response of EAS arrays to low energy events is compromised at large zenith angles. Still, as shown in Table~\ref{undetected} and in~\S\ref{sec:extra-compa}, four of the corresponding upper limits are below the extrapolation of the corresponding 3FHL spectrum.

IC~310 is a target that is both nearby and culminates at an adequate zenith angle, but remained undetected in this HAWC analysis. The upper limit set in the photon flux is close to a third of the LAT extrapolation, and the HAWC upper limit in the 0.5-2.0~TeV band is 2.5 lower than the one in the 3FHL catalog. We note that while the 3FHL catalog reports a relatively low variability index, $V_{\rm bayes}=2$, this source is known to display extreme variability in the VHE range, on timescales as low as five~minutes, challenging models and severely constraining the emission region to scales smaller than its event horizon~\citep{2014Sci...346.1080A}.

In addition to the preselected targets, we point here to two additional sources of intrinsic interest: 3C~264 and NGC~1068. 3C~264 is a radio galaxy hosted by the elliptical galaxy NGC~3862, at a distance of about 90~Mpc. A compact radio source powers a relativistic jet imaged in radio and in the optical~\citep{2004A&A...415..905L,1993ApJ...402L..37C}. 3C~264 shows in the LAT data up to a highest energy photon of 97~GeV. It was listed as a TeV=N source in the 3FHL catalog, and considered as such in this analysis. However, it was later detected in the VHE range by the VERITAS Collaboration~\citep{2018ATel11436....1M}. The photon flux measured by VERITAS above 300~GeV indicates that this object should be too faint for HAWC. Even if the spectral index measured by LAT ($\alpha=1.65$) were to continue in the TeV range, its extrapolation would be $N(>0.5\,{\rm TeV})\lesssim 0.6\times 10^{-12}\,\rm cm^{-2}s^{-1}$. The HAWC data shows a $+1.9\sigma$ excess, statistically consistent with such a low flux and providing an upper limit  $N(>0.5\,{\rm TeV}) < 1.42\times 10^{-12}\,\rm cm^{-2}s^{-1}$.

NGC~1068 has become a source of interest due to its coincidence with a hotspot in the IceCube all-sky map~\citep{2020PhRvL.124e1103A}. While two other starburst galaxies, NGC~253 and M82, have been detected in the VHE range as faint sources, with fluxes 1\% of the Crab Nebula~\citep{2016CRPhy..17..585O}, NGC~1068 remains undetected and an unlikely candidate for HAWC detection. It is a weak 5.3$\sigma$ detection in the 3FHL catalog, with practically no signal above 20~GeV and a rather steep spectral index \mbox{$\alpha=3.8\pm 1.0$}.  The HAWC data has a $+1.2\sigma$ excess at the location of NGC~1068, for an upper limit normalization $K_{2\sigma}= 6.46\times 10^{-13}\,\rm TeV^{-1} cm^{-2}s^{-1}$ at 1~TeV. The quasi-differential HAWC limit in the common energy band, $N(0.5-2.0~{\rm TeV})\leq 2.32\times 10^{-12}\,\rm cm^{-2}s^{-1}$, is a factor of eight lower than the respective LAT limit. 

\section{Summary \label{sec:summary}}
The HAWC Gamma-Ray Observatory has performed an extensive follow-up survey of known GeV $\gamma$-ray emitting active galaxies at TeV energies. We investigated all AGN in the 3FHL catalog with a redshift lower than 0.3 and transiting within $40^\circ$ of latitude $19^\circ\,\rm N$, the HAWC zenith, searching for TeV $\gamma$-ray emission averaged over a 4.5~year period. The HAWC data show clear signals from Mkn~421 and Mkn~501, from which we quantified their long-term averaged spectra. In addition, we obtained evidence for TeV emission from the radiogalaxy M87 and the BL Lac objects VER~J0521+211 and 1ES~1215+303. The fluxes estimated for these sources are compatible with values previously reported for mid or low activity states. When excluding Mkn~421 and Mkn~501, we find collective evidence for long-term averaged TeV emission from radiogalaxies and BL Lac objects with a p-value $\sim 1\%$, and for known VHE emitters (TeV=P) with p-value $\sim 10^{-5}$. No evidence was found for other source classes or for LAT sources not previously reported in the VHE range. 

Upper limits were set for the whole sample assuming intrinsic power-law spectra of index 2.5 attenuated by the interaction of $\gamma$ rays with extragalactic background radiation. These limits are dependent on the declination and redshift of the sources, confirming a redshift attenuation of exponential scale $z_h \simeq 0.1$ for HAWC. Bounds on observed photon fluxes in three energy intervals (0.5-2.0)~TeV, (2.0-8.0)~TeV and (8.0-32.0)~TeV are also given. HAWC measurements were compared with the mostly non-contemporaneous 3FHL catalog long-term data and with specific IACT observations. 

As the exposure of HAWC continues to deepen, the increased sensitivity will allow to perform deeper searches of extragalactic sources. Long-term variability is an area of opportunity for ground based EAS $\gamma$-ray observatories not explored in this paper. Analyses of multi-year AGN light curves are underway in order to expand the investigation presented here to the time regime. The continuous and improved operation of HAWC is leading to a better understanding of EAS arrays, and in particular of water Cherenkov detectors. New analysis tools to improve the sub-TeV sensitivity of HAWC have been developed, and are now been implemented. These will provide improved energy response and the reduction of noise at low energies, two requirements for a deeper access to the extragalactic sky. These upgrades will allow HAWC to build on the analysis presented here, laying the path to future TeV survey instruments, in both the Northern and Southern Hemispheres.

\acknowledgments

We acknowledge the support from: the US National Science Foundation (NSF); the US Department of Energy Office of High-Energy Physics; the Laboratory Directed Research and Development (LDRD) program of Los Alamos National Laboratory; Consejo Nacional de Ciencia y Tecnolog\'ia (CONACyT), M\'exico, grants 271051, 232656, 260378, 179588, 254964, 258865, 243290, 132197, A1-S-46288, A1-S-22784, C\'atedras 873, 1563, 341, 323, Red HAWC, M\'exico; DGAPA-UNAM grants IG100317, IG101320, IN111315, IN111716-3, IN111419, IA102019, IN112218; VIEP-BUAP; PIFI 2012, 2013, PROFOCIE 2014, 2015; the University of Wisconsin Alumni Research Foundation;  the Institute of Geophysics, Planetary Physics, and Signatures at Los Alamos National Laboratory; Polish Science Centre grant, DEC-2017/27/B/ST9/02272; Coordinaci\'on de la Investigaci\'on Cient\'ifica de la Universidad Michoacana; the Royal Society - Newton Advanced Fellowship 180385; la Generalitat Valenciana, grant CIDEGENT/2018/034. Chulalongkorn University's CUniverse (CUAASC) grant. Thanks to Scott Delay, Luciano D\'iaz and Eduardo Murrieta for technical support.

This work used data from the Fermi Science Support Center and the TeVCat online source catalog (http://tevcat.uchicago.edu). It also made use of the SIMBAD, NED (NASA/IPAC Extragalactic Database), ADS, and SDSS databases.

%

\vspace{5mm}
\facilities{The High Altitude Water Cherenkov (HAWC) Gamma-Ray Observatory}

\bibliography{agnsurvey_hawc}{}
\bibliographystyle{aasjournal}


\end{document}